\documentclass{jfm}

\usepackage{graphicx}
\usepackage{newtxtext}
\usepackage{newtxmath}
\usepackage{natbib}
\usepackage{hyperref}
\usepackage{enumerate}
\usepackage{enumitem}
\usepackage{rotating}
\usepackage[usenames, dvipsnames]{xcolor}
\hypersetup{
    colorlinks = true,
linkcolor=MidnightBlue,
    urlcolor   = blue,
    citecolor  = MidnightBlue,
}
\usepackage{pdflscape}
\usepackage{afterpage}
\usepackage{flafter}
\usepackage{rotating}
\usepackage{fancyhdr}
\fancypagestyle{lscape}{% 
\fancyhf{} % clear all header and footer fields 
\fancyfoot[LE]{}
\fancyfoot[LO] {}}

\newcommand*{\ep}{\epsilon}

\renewcommand*{\i}{\mathrm{i}}

\newcommand{\RomanNumeralCaps}[1]
\linenumbers
\graphicspath{ {images/} }

\shorttitle{Viscous gravity-capillary waves}
\shortauthor{Shelton, Milewski, and Trinh}

\title{Time-dependent nonlinear gravity-capillary surface waves with viscous dissipation and wind forcing}

\author{Josh Shelton\aff{1,2}\corresp{\email{josh.shelton@st-andrews.ac.uk}}, Paul Milewski\aff{3}
 \and Philippe H. Trinh\aff{2}}

\affiliation{
\aff{1}School of Mathematics and Statistics, University of St Andrews, St Andrews, KY16 9SS, UK
\aff{2}Department of Mathematical Sciences, University of Bath, Bath BA2 7AY, UK
\aff{3}Department of Mathematics, Penn State University, Pennsylvania 16802, USA}

\date{\today~[Draft]}

\AtBeginDocument{}
\begin{document}
\maketitle
 \hrule width \linewidth height 0.6pt
 \vspace{2.5mm}
\begin{abstract}
We develop a time-dependent conformal method to study the effect of viscosity on steep surface waves. When the effect of surface tension is included, numerical solutions are found that contain highly oscillatory parasitic capillary ripples. These small amplitude ripples are associated with the high curvature at the crest of the underlying viscous-gravity wave, and display asymmetry about the wave crest. 
Previous inviscid studies of steep surface waves have calculated intricate bifurcation structures that appear for small surface tension.
We show numerically that viscosity suppresses these. While the discrete solution branches still appear, they collapse to form a single smooth branch in limit of small surface tension. These solutions are shown to be temporally stable, both to small superharmonic perturbations in a linear stability analysis, and to some larger amplitude perturbations in different initial-value problems. Our work provides a convenient method for the numerical computation and analysis of water waves with viscosity, without evaluating the free-boundary problem for the full Navier-Stokes equations which becomes increasingly challenging at larger Reynolds numbers.
\end{abstract}
 \vspace{-0.5mm}
 \hrule width \linewidth height 0.6pt
  \vspace{-4mm}
 
\section{Introduction}\label{sec:intro}

When surface tension is included in the classical formulation of a travelling nonlinear gravity wave, solutions are seen to exhibit highly oscillatory parasitic capillary ripples. These are small scale ripples which travel with the same speed as the underlying gravity wave. In most previous numerical and asymptotic studies \citep{schwartz_1979,shelton2021structure,shelton2022exponential}, this phenomena has been studied with the use of potential flow theory, in which the effect of viscosity is neglected. This results in symmetric solutions, in which the capillary ripples are observed across the entire surface wave profile. This is in contrast to experimental observations \citep{ebuchi_1987}, in which the parasitic ripples are focused on the forward face of the travelling wave. This disparity between the surface profiles viewed experimentally, and those from inviscid theory, is usually attributed to the effect of fluid viscosity \citep{dosaev2021physical}. 

It is of significant interest to develop simplified models for gravity-capillary waves which include viscosity, but without resorting to solving the full Navier-Stokes equations. To this aim, the principal question is whether it is possible to combine the simplicity of the classical potential flow framework for an inviscid water-wave, with effective conditions that govern and model viscous effects near key regions. Note that the viscous terms in the Navier-Stokes equations are identically zero for a velocity derived from a potential. In the limit of small viscosity, analytical progress is tractable through the study of a viscous boundary layer near to the fluid surface \citep{longuet1992capillary}. Previous authors, such as \cite{ruvinsky1991numerical} and \cite{fedorov1998nonlinear}, have applied these techniques to investigate the effect of viscosity on steep gravity-capillary waves. Both of these investigations produced solutions with asymmetric parasitic capillary ripples, which closely resembled previous experimental observations. In the work by \cite{ruvinsky1991numerical}, kinematic and dynamic boundary conditions were presented, where viscosity feeds into the kinematic condition through a nonlocal condition involving time integration of the velocity potential -- a step that was subsequently simplified into a local condition by \cite{dias2008theory}. Approximations valid in the limit of small amplitude were then made to determine explicit equations for the Fourier coefficients of the solution, which also yielded the damping rate as a function of the small amplitude parameter. A steady formulation of this problem has also been developed by \cite{fedorov1998nonlinear} using viscous boundary layer approximations. With the addition of a surface pressure forcing (modelling the effect of wind), steadily travelling solutions were calculated numerically. This model was subsequently used by \cite{melville2015equilibrium} to demonstrate that in the ocean, parasitic capillary ripples, rather than the main gravity wave, can be responsible for the majority of the viscous damping required to offset the growth of the overall wave due to the effect of wind. We note
that the earliest work investigating the effect of viscosity on steep waves with parasitic capillary ripples was by \cite{longuet1963generation}; however these results compared poorly with the experimental observations by \cite{perlin1993parasitic}. This theory would later be improved upon and updated by \cite{longuet_1995}. While they proposed many groundbreaking ideas, the above works are often challenging to interpret mathematically on account of the number of approximations made.

In this work, we shall study the formulation proposed by \cite{dias2008theory} which incorporates viscosity in the surface boundary conditions of a potential flow boundary-value problem. In this model, the free-surface kinematic and dynamic boundary conditions are given by
\begin{equation}\left. \quad
\begin{aligned}\label{eq:intro}
\phi_t +\frac{1}{2}(\phi_x^2+\phi_y^2) + \frac{\zeta}{F^2}-\frac{B}{F^2} \kappa +\frac{P}{F^2}\zeta_{x}+ \frac{2}{Re}\phi_{yy}=0,\\
\zeta_t  = \phi_y - \phi_x \zeta_x + \frac{2}{Re} \zeta_{xx},
\end{aligned}\quad \right\}
\end{equation}
evaluated at the unknown free surface, $y=\zeta(x,t)$. In boundary conditions \eqref{eq:intro}, $\phi(x, t)$ is the velocity potential which satisfies Laplace's equation within the domain $-1/2 \leq x \leq 1/2$, $-\infty < y \leq \zeta(x,t)$. 
The formulation includes the effect of surface tension through the Bond number, $B$, and curvature, $\kappa=\zeta_{xx}/(1+\zeta_x^2)^{3/2}$, the effect of fluid viscosity through the Reynolds number, $Re$, and surface wind forcing which depends on the wave slope, $\zeta_{x}$. The wind forcing is required in order to obtain steadily travelling solutions in the presence of viscous dissipation.  Full details of the mathematical formulation are given later in \S\,\ref{sec:formulation} and particularly in equations \eqref{eq:main1}-\eqref{eq:main4}. Finding solutions to this formulation is difficult on account of the moving domain, $-\infty \leq y \leq \zeta(x,t)$, and the dependence of the solution on three independent variables, $x$, $y$, and $t$. 

Our work focuses on the development of a time-dependent conformal method that allows for the convenient numerical evaluation of a two-dimensional potential flow problem with viscosity. The main benefit of this conformal method is that the original two-dimensional domain, which is bounded by a moving surface wave, is reduced to a fixed one-dimensional domain for the time-dependent free surface. We apply the conformal mapping techniques of \cite{dyachenko1996nonlinear} to system \eqref{eq:intro} to derive time-dependent equations that govern each variable on the free surface. These will depend only on one spatial coordinate, the conformal variable $\xi$, which parametrises the free surface. This conformal method preserves all features of the original boundary-value problem, and no further assumptions are made following the introduction of boundary conditions \eqref{eq:intro}.

\begin{figure}
\centering
\includegraphics[scale=1]{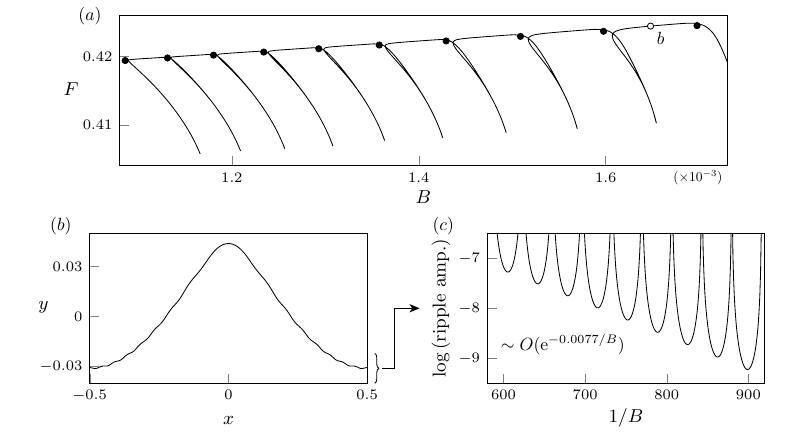}
\caption{\label{fig:sheltonetal} Numerical results for steadily propagating inviscid gravity-capillary waves calculated by \cite{shelton2021structure}. $(a)$ The location of solutions is shown in the $(B,F)$-plane, where the Bond number $B$ and Froude number $F$ are nondimensional parameters defined in \eqref{eq:nondimparams}. Black circles show the parameter values given from the failure the nonlinear solvability condition derived by \cite{shelton2022exponential}. $(b)$ A typical solution containing oscillatory capillary ripples, which has $B=0.001648$ and $F=0.4245$. $(c)$ Semi-log plot showing the exponentially-small amplitude of the capillary ripples for each solution branch in $(a)$.}
\end{figure}

 In previous investigations of inviscid surface waves, intricate bifurcation structures have emerged in the small surface tension limit \citep{champneys_2002,shelton2021structure,shelton2023structure}, an example of which is shown for steadily travelling gravity-capillary waves in figure~\ref{fig:sheltonetal}$(a)$. These consist of a countably infinite number of solution branches in the $(B,F)-$plane that manifest for fixed amplitude under the limit of small surface tension. Across each of the branches in figure~\ref{fig:sheltonetal}$(a)$, the wavenumber of the parasitic capillary ripples increases by one. Figure~\ref{fig:sheltonetal}$(b)$ shows an example of a travelling surface wave obtained in this inviscid framework, in which the capillary ripples are symmetric about the wave crest. The amplitude of these parasitic capillary ripples was measured by \cite{shelton2021structure} to be exponentially small as $B \to 0$. Asymptotic solutions for these were later obtained by \cite{shelton2022exponential} using beyond-all-order asymptotics, in which the capillary ripples were produced by the Stokes phenomenon across Stokes lines associated with the high crest curvature of the leading-order nonlinear gravity wave. This also produced an asymptotic solvability condition for when perturbation solutions do and do not exist -- values of nonexistence are shown with black circles in figure~\ref{fig:sheltonetal}$(a)$. The exponential scaling of the capillary ripple amplitude is shown in the semi-log plot of figure~\ref{fig:sheltonetal}$(c)$.

We use the nonlinear formulation \eqref{eq:intro} to investigate the effect of small capillarity on viscous gravity-capillary waves. First, we study steadily travelling solutions, for which surface wind forcing is introduced in the kinematic boundary condition to counteract the energy decay induced by viscous dissipation. We demonstrate numerically that in the presence of viscosity, the discrete branching structure obtained in the inviscid regime (figure~\ref{fig:sheltonetal}$(a)$) does not persist below a certain value of the surface tension. While discrete branches are observed for larger values of the surface tension, they close off such that a single smooth branch exists in the limit of zero surface tension. This is observed numerically both when the viscosity is fixed, and when distinguished limits are chosen in which the viscosity decays alongside the surface tension in an algebraic manner.

Secondly, we study the temporal stability of our steady parasitic solutions. These are shown to be superharmonically stable through a linear stability analysis, in which an eigenvalue problem for small perturbations is studied and these are shown to decay in time. A nonlinear stability analysis is then considered with an initial-value problem which implements our time-dependent formulation.
This allows us to comment on the global `attractiveness' of the steadily travelling solutions, and convergence is observed when starting from both a large amplitude initial condition, and a small amplitude initial condition.
In this latter case, the surface wind forcing initially dominates which causes the wave amplitude to increase, before eventually balancing with viscous dissipation as the target solution is converged upon.

\subsection{Outline of our paper}
We begin in \S\,\ref{sec:formulation} by formulating the two-dimensional boundary-value problem that models nonlinear surface waves which travel upon a viscous fluid. Our conformal mapping of the unsteady problem, for which full details are presented in Appendix~\ref{App:timemap}, yields a one-dimensional formulation for the free surface. This is given in \S\,{\ref{sec:confmaptime}} for unsteady solutions, and \S\,{\ref{sec:confmap}} for steadily travelling solutions. Numerical solutions to the steady formulation are obtained in \S\,\ref{sec:steadyresults}, with an emphasis on detecting the bifurcation structure that emerges for small surface tension and viscosity. The unsteady formulation is implemented numerically in \S\,{\ref{sec:timeresults}}, where we perform a linear and nonlinear stability analysis to observe the temporal stability of the steady solutions when starting from different initial conditions.

\section{Mathematical formulation}\label{sec:formulation}
We consider the time evolution of a nonlinear surface wave subject to small viscous effects within a two-dimensional fluid extending to infinite depth. The travelling wave is assumed to be periodic in the direction of propagation, and is subject to the effects of both gravity and surface tension. The boundary-value problem describing this is specified on the fluid domain $-1/2 \leq x \leq  1/2$ and $-\infty < y \leq \zeta(x,t)$, with the velocity potential $\phi(x,y,t)$ and the free surface $y=\zeta(x,t)$ as solutions. After nondimensionalisation, and moving into a co-moving frame of unit speed through a Galilean transformation, the boundary-value problem is given by
\begin{subequations}\label{eq:main}
\begin{align}\label{eq:main1}
\phi_t-\phi_{x}  +\frac{1}{2}(\phi_x^2+\phi_y^2) + \frac{\zeta}{F^2}-\frac{B}{F^2} \kappa +\frac{P}{F^2}\zeta_x+ \frac{2}{Re}\phi_{yy}=0& \qquad \text{at} ~~~ y=\zeta(x,t),\\
\label{eq:main2}
\zeta_t  = \phi_y+\zeta_x(1-\phi_x) + \frac{2}{Re} \zeta_{xx}& \qquad \text{at} ~~~ y=\zeta(x,t),\\
\label{eq:main3}
\phi_{xx} +\phi_{yy}=0& \qquad \text{for} ~~ y \leq \zeta(x,t),\\
\label{eq:main4}
\phi_{x} \to 0, \quad \phi_{y} \to 0& \qquad \text{as} ~~ y \to -\infty.
\end{align}
\end{subequations}
This system comprises of kinematic and dynamic boundary conditions \eqref{eq:main1} and \eqref{eq:main2} on the surface, Laplace's equation \eqref{eq:main3} within the fluid, and the decay of motion in the deep water limit \eqref{eq:main4}. The effect of viscosity appears in the boundary conditions \eqref{eq:main1} and \eqref{eq:main2}; these model effects were proposed by \cite{dias2008theory}. Note also the inclusion of the surface wind forcing, $P \zeta_{x}/F^2$ in \eqref{eq:main1}. 
We will refer to $P$ as the nondimensional wind strength.
Note that in the inviscid formulation with $1/Re=0$ and $P=0$, the wave energy, which measures the effects of kinetic, capillary, and gravitational potential energies, is a time-conserved quantity of the unsteady formulation. However, when $1/Re \neq 0$ and $P=0$, dissipation renders the energy a decreasing function of time for solutions to formulation \eqref{eq:main}, which we demonstrate in Appendix~\ref{App:energydecay}. Thus, steadily travelling waves will not exist in this viscous formulation in the absence of wind forcing.
This is the reason why we include both the effects of viscous dissipation and wind forcing in this work, in order to study and classify steadily travelling solutions, as well as analyse their stability.
This model for wind forcing is discussed in more detail in \S\,\ref{sec:pressure}.

Four nondimensional constants appear in system \eqref{eq:main} above. These are the Froude number, $F$, the Bond number, $B$, the Reynolds number, $Re$, and the nondimensional wind strength, $P$, defined by
\begin{equation}\label{eq:nondimparams}
F= \frac{c}{\sqrt{g \lambda}}, \qquad B=\frac{\sigma}{\rho g \lambda^2}, \qquad Re=\frac{c \lambda}{\nu}, \qquad P=\frac{p}{g \lambda}.
\end{equation}
Here, $c$ is the speed of the travelling frame (which for steady solutions is the wavespeed), $g$ is the constant acceleration due to gravity, $\lambda$ is the wavelength, $\sigma$ is the constant coefficient of surface tension, $\rho$ is the fluid density, $\nu$ is the kinematic viscosity of the fluid, and $p$ is the amplitude of the physical wind forcing. Note that we have nondimensionalised length scales and the free surface elevation with respect to $\lambda$, the velocity potential with respect to $c \lambda$, and time with respect to $\lambda/c$.

Crucially, we note that, when considering the full effects of both nonlinearity and viscosity, there is no known explicit formulation of the kinematic and dynamic boundary conditions, which can be written in a form analogous to \eqref{eq:main1} and \eqref{eq:main2} for the velocity potential and surface height. In such cases, it is necessary to solve the full Navier-Stokes equations within the unknown fluid domain; this has been performed numerically by {\itshape e.g.} \cite{hung2009formation} for the case of time-dependent nonlinear viscous gravity-capillary waves. However, explicit kinematic and dynamic boundary conditions for the velocity potential, $\phi$, and free-surface, $\zeta$, can be derived in two cases:
\begin{enumerate}[label=(\roman*),leftmargin=*, align = left, labelsep=\parindent, topsep=3pt, itemsep=2pt,itemindent=0pt ]
\item Inviscid flows, for which the kinematic and dynamic boundary conditions for gravity-capillary waves emerge, the former from the classical Bernoulli equation (cf. \citealt{vb_book}). These are the same as equations \eqref{eq:main} but with the viscous terms, which contain the Reynolds number, $Re$, removed.
\item Linear flows with non-zero viscosity: here, the kinematic and dynamic boundary conditions may be derived from the linearised 2D Navier-Stokes equations. This is performed by decomposing the velocity field into irrotational and solenoidal components, for which the linearised forms of \eqref{eq:main1} and \eqref{eq:main2} were derived by \cite{dias2008theory}. These equations have subsequently been used to study viscous effects on a variety of free-surface formulations, such as for Faraday pilot waves in bouncing fluid droplets \citep{milewski2015faraday,blanchette2016modeling}, three dimensional solitary waves with forcing \citep{wang2012dynamics}, and deriving dissipation rates for ocean swell \citep{henderson2013role}.
\end{enumerate}

Thus, the equations that we use throughout this work are obtained by combining the viscous term found from (ii) above with the nonlinear inviscid boundary conditions from (i), yielding \eqref{eq:main1} and \eqref{eq:main2}. We note that the `correct' nonlinear generalisation of the term $\phi_{yy}$ in kinematic condition \eqref{eq:main1} is $\phi_{nn}$, where $n$ is the normal vector to the free surface, which was originally derived by \cite{ruvinsky1991numerical}. There is no known correct nonlinear generalisation of the term $\zeta_{xx}$ in \eqref{eq:main2}. These model equations were proposed by \cite{dias2008theory}, and one aim of our current work is to develop a time-dependent conformal map that efficiently solves the nonlinear version of this problem. This methodology is then used to study the effect that viscosity plays in the parasitic capillary ripples present on steep gravity waves, their associated bifurcation structure, and temporal stability. We are particularly interested in the small surface tension limit on account of the intricate bifurcation structures that exist in this singular regime.
We note that these nonlinear equations have previously been applied to study dissipative effects on solitons by \cite{brunetti2014nonlinear} and \cite{liao2023modified}, both of whom derived forced NLS equations asymptotically.

\subsection{Choice of wind forcing}\label{sec:pressure}
To obtain steady travelling solutions, we balance viscous dissipation with the addition of a wind forcing term in Bernoulli's equation \eqref{eq:Tmapphievolve}. 
The surface wind forcing we consider is that developed by \cite{jeffreys1925formation} as a model for linear wave growth induced by wind forcing. This results in the addition of the term $p\hat{\zeta}_{\hat{x}}$ into the dimensional kinematic condition, which after nondimensionalisation becomes $P \zeta_{x}/F^2$ in \eqref{eq:main2}. Here, $p$ is the dimensional wind strength and $P=p/(g \lambda)$ is the nondimensional wind strength.
This is known as a sheltering model; for a wave travelling from left to right with $P>0$ it adds energy to the rear face of the wave, and removes energy from the forward face.
In our numerical search for steadily travelling solutions in \S\,\ref{sec:steadyresults}, the constant $P$ is treated as an unknown of the formulation, and a unique value is obtained by solving an underdetermined system of equations.

In the study by \cite{fedorov1998nonlinear}, who included forcing in the kinematic condition, the term $P\cos{(2 \pi x)}$ was used to mimic the effect of wind. Federov and Melville found that when all other parameters were fixed, a range of $P$ values were permitted, corresponding to different phase shifts between their fixed wind forcing and unknown solution profile. This complication occurred as a result of their choice of wind forcing breaking the translational invariance of the system.
In contrast, the Jeffreys model for wind forcing that we use in this work both retains the translational invariance of the system and is a justifiable model for the shear stress induced by wind forcing.
Since a unique value of $P$ is selected as an unknown to the steady solutions, the resultant bifurcation space is simpler than that investigated by \cite{fedorov1998nonlinear}, which allows for an easier study of solution branches.

We note that more accurate models exist to capture the influence of wind on surface water waves. The monograph by \cite{janssen2004interaction} details elements incorporated in these, such as the transfer of energy between coupled air and water layers, and the consideration of turbulent effects in the air.
For instance, the formulation from which \cite{miles1957generation} developed his theory of linear wave generation involved inviscid and incompressible air and water layers coupled by an interfacial stress condition. The water layer was irrotational, and the consideration of rotational effects in the air layer gave rise to a critical-layer problem governed by the inviscid Orr-Sommerfeld equation for linear disturbances to the wave surface.

\subsection{The time-dependent conformal mapping}\label{sec:confmaptime}
Evolving the solutions $\zeta(x,t)$ and $\phi(x,y,t)$ to the boundary-value problem \eqref{eq:main} in time is difficult. This is both due to the physical $(x,y)$ domain being two dimensional, and the boundary conditions being imposed upon the free surface, which itself can change in time. In this section, we present a time-dependent mapping of system \eqref{eq:main}, under which the flow domain $-1/2 \leq x \leq  1/2$ and $-\infty < y \leq \zeta(x,t)$ is mapped to the lower-half $(\xi,\eta)$ plane. Upon evaluating the solutions at the fluid surface, $\eta=0$, a one-dimensional surface formulation emerges, parameterised by $\xi$ and $t$. This conformal method was originally developed by \cite{dyachenko1996nonlinear} and \cite{choi1999exact} for nonlinear gravity capillary waves.

In writing $x=x(\xi,\eta,t)$ and $y=y(\xi,\eta,t)$, the surface solutions $X$, $Y$, $\Phi$, and $\Psi$ are defined by
\begin{equation}\left. \quad
\begin{aligned}\label{eq:Tmapvariables}
X(\xi,t)=x(\xi,0,t),& \qquad \Phi(\xi,t)=\phi(x(\xi,0,t),y(\xi,0,t),t),\\
Y(\xi,t)=\zeta(x(\xi,0,t),t),& \qquad \Psi(\xi,t) = \psi(x(\xi,0,t),y(\xi,0,t),t).
\end{aligned}\quad \right\}
\end{equation}
Here, the streamfunction $\psi$ is the harmonic conjugate of the velocity potential $\phi$. We then derive time-evolution equations for the surface solutions \eqref{eq:Tmapvariables}, for which full details are presented in Appendix~\ref{App:timemap}. We obtain evolution equations for the surface height $Y$,
\begin{subequations}\label{eq:Tmapevolveboth}
\begin{equation}\label{eq:Tmapyevolve}
Y_t=\frac{X_{\xi}(Y_{\xi}-\Psi_{\xi})}{J}+\frac{2}{Re}\frac{X_{\xi} Y_{\xi \xi} - Y_{\xi} X_{\xi \xi}}{X_{\xi}J}-Y_{\xi} \mathscr{H} \bigg[\frac{Y_{\xi}-\Psi_{\xi}}{J} +\frac{2}{Re}\frac{X_{\xi}Y_{\xi \xi}-Y_{\xi}X_{\xi\xi}}{X_{\xi}^2 J}\bigg],
\end{equation}
and the surface velocity potential $\Phi$,
\begin{equation}\label{eq:Tmapphievolve}
\begin{aligned}
\Phi_t=&\frac{\Psi_{\xi}^2-\Phi_{\xi}^2}{2J} - \frac{Y}{F^2}-\frac{P}{F^2}\frac{Y_{\xi}}{X_{\xi}} + \frac{B \kappa}{F^2}+\frac{X_{\xi}\Phi_{\xi}}{J}-\Phi_{\xi}\mathscr{H} \bigg[\frac{Y_{\xi}-\Psi_{\xi}}{J} +\frac{2}{Re}\frac{X_{\xi}Y_{\xi \xi}-Y_{\xi}X_{\xi\xi}}{X_{\xi}^2 J}\bigg] \\
&+\frac{2}{Re}\bigg[\frac{(Y_{\xi} X_{\xi \xi} - X_{\xi} Y_{\xi \xi})}{X^2_{\xi}J}\Psi_{\xi}
+\frac{Y_{\xi}X_{\xi \xi}(Y_{\xi}^2-3X_{\xi}^2)+X_{\xi}Y_{\xi\xi}(X_{\xi}^2-3Y_{\xi}^2)}{J^3}\Psi_{\xi}\\
&+\frac{X_{\xi}X_{\xi \xi}(3Y_{\xi}^2-X_{\xi}^2)+Y_{\xi}Y_{\xi \xi}(Y_{\xi}^2-3X_{\xi}^2)}{J^3}\Phi_{\xi} +\frac{X_{\xi}^2-Y_{\xi}^2}{J^2}\Phi_{\xi \xi}+\frac{2X_{\xi}Y_{\xi}}{J^2}\Psi_{\xi \xi}\bigg],
\end{aligned}
\end{equation}
where $X_{\xi}$ and $\Psi$ are known from the harmonic relations
\refstepcounter{equation}\label{eq:harmonicrelationsC}
\refstepcounter{equation}\label{eq:harmonicrelationsD}
\begin{equation}
X_{\xi}=1-\mathscr{H}[Y_{\xi}] \quad \text{and} \quad \Psi = \mathscr{H}[\Phi].
\tag{\ref*{eq:harmonicrelationsC},$d$}
\end{equation}
\end{subequations}
In system \eqref{eq:Tmapevolveboth} above, $J=X_{\xi}^2+Y_{\xi}^2$ is the Jacobian of the mapping, $\kappa=(X_{\xi}Y_{\xi \xi}-Y_{\xi}X_{\xi \xi})/J^{3/2}$ is the surface curvature, and $\mathscr{H}[Y]=\int_{-1/2}^{1/2}Y(\xi^{'}) \cot{[\pi(\xi^{'}-\xi)]}\mathrm{d}\xi^{'}$ is the periodic Hilbert transform.
Given an initial condition for $Y(\xi,0)$, $\Phi(\xi,0)$, and values of $B$, $F$, $Re$, and $P$, equations \eqref{eq:Tmapevolveboth} may be used to evolve the solutions in time. In \S\,\ref{sec:timeresults}, we use this scheme to study the stability of steadily travelling solutions with surface wind forcing, when viewed as convergence in a time-evolution problem.

\subsection{Conformal mapping for steadily travelling waves}\label{sec:confmap}
We now present integro-differential equations, depending only on the conformal domain $\xi$, that are satisfied by steady solutions of the mapped system \eqref{eq:Tmapevolveboth}. These are given by
\begin{subequations}\label{eq:steadyequations}
\begin{equation}
\begin{aligned}\label{eq:steadyequationsa}
&\frac{\Phi_{\xi}^2+\Psi_{\xi}^2}{2J}-\frac{(X_{\xi}\Phi_{\xi}+Y_{\xi} \Psi_{\xi})}{J} + \frac{Y}{F^2} +\frac{P}{F^2}\frac{Y_{\xi}}{X_{\xi}} -\frac{B}{F^2}\frac{(X_{\xi} Y_{\xi \xi} - Y_{\xi} X_{\xi \xi})}{J^{3/2}} \\
&+\frac{2}{Re}\bigg[\frac{(Y_{\xi}^2-X_{\xi}^2)\Phi_{\xi \xi}-2 X_{\xi}Y_{\xi}\Psi_{\xi \xi}}{J^2} +\frac{\Phi_{\xi}}{J^3}\bigg( X_{\xi \xi} X_{\xi}(X_{\xi}^2-3Y_{\xi}^2)+Y_{\xi \xi}Y_{\xi}(3X_{\xi}^2-Y_{\xi}^2)\bigg)\\
&\qquad \quad  +\frac{\Psi_{\xi}}{J^3}\bigg(X_{\xi \xi} Y_{\xi}(3X_{\xi}^2-Y_{\xi}^2)+Y_{\xi \xi} X_{\xi}(3Y_{\xi}^2-X_{\xi}^2)\bigg)\bigg] =0,
\end{aligned}
\end{equation}
\begin{equation}\label{eq:steadyequationsb}
\Psi_{\xi} = Y_{\xi} +\frac{2}{Re} \frac{X_{\xi} Y_{\xi \xi} - Y_{\xi} X_{\xi \xi}}{X_{\xi}^2},
\end{equation}
together with harmonic relations 
\refstepcounter{equation}\label{eq:steadyharmonicrelationsC}
\refstepcounter{equation}\label{eq:steadyharmonicrelationsD}
\begin{equation}
X_{\xi}=1-\mathscr{H}[Y_{\xi}] \quad \text{and} \quad \Phi_{\xi}=-\mathscr{H}[\Psi_{\xi}],
\tag{\ref*{eq:steadyharmonicrelationsC},$d$}
\end{equation}
\end{subequations}
where $\mathscr{H}$ is the periodic Hilbert transform. This yields four equations for the four unknown functions $X$, $Y$, $\Phi$, and $\Psi$. 

There are two ways to derive system \eqref{eq:steadyequations}. First, one may consider steady solutions of the original boundary-value problem \eqref{eq:main}, for which a steady conformal mapping analogous to that presented in Appendix~\ref{App:timemap} yields equations \eqref{eq:steadyequationsa} and \eqref{eq:steadyequationsb}. Alternatively, one may consider steady solutions of the time-dependent conformal system \eqref{eq:Tmapevolveboth}, which after simplification also yields \eqref{eq:steadyequationsa} and \eqref{eq:steadyequationsb}.

When solving \eqref{eq:steadyequations} numerically, we will typically fix $B$, $Re$, and the wave energy $\mathscr{E}$, defined by
\begin{subequations}
\begin{equation}\label{eq:energy}
\mathscr{E}=\frac{1}{E_{\text{hw}}}\int_{-1/2}^{1/2} \bigg[ \underbrace{\frac{F^2}{2}\Psi_{\xi} \Phi}_{\text{kinetic}}+\underbrace{\vphantom{\frac{F^2}{2}}B(\sqrt{J}-X_{\xi})}_{\text{capillary}}+\underbrace{\frac{1}{2}Y^2X_{\xi}}_{\text{gravitational}}\bigg] \mathrm{d}\xi,
\end{equation}
for which $F$ and $P$ are obtained as unknowns of the problem. This is performed in \S\,\ref{sec:steadyresults} to calculate nonlinear solutions for small values of the surface tension. The numerical method used to solve system \eqref{eq:steadyequations} is given in \S\,\ref{sec:numericalsteady}. In \eqref{eq:energy} above, we have normalised with respect to the energy of the highest inviscid Stokes wave, $E_{\text{hw}}=0.00184$, computed to three significant digits. The steady solutions calculated in this paper will have $\mathscr{E}=0.4$, which is chosen to compare with previous inviscid works \citep{shelton2021structure,shelton2022exponential} studying the classes of parasitic solutions that emerge in the small surface tension limit.
Note that the energy is not a conserved quantity in this formulation, due to the presence of wind forcing and dissipation. An expression for the rate of change of energy in the absence of the surface wind forcing is derived in Appendix~\ref{App:energydecay}, and this is shown to be inversely proportional to $Re$.
Note also that the steady solutions only provide information of $\Phi_{\xi}$, whereas evaluation of \eqref{eq:energy} requires $\Phi$, which we obtain by integration. The constant of integration for $\Phi$ is explicitly determined from the condition
\begin{equation}\label{eq:phicond}
\int_{-1/2}^{1/2} \Phi X_{\xi} \mathrm{d}\xi =0.
\end{equation}
\end{subequations}
Condition \eqref{eq:phicond} is equivalent to writing $\int_{-1/2}^{1/2}\phi(x,\zeta(x)) \mathrm{d}x=0$ in conformal variables.

\subsection{Linear solutions for the surface and the internal vorticity field}
In the \cite{dias2008theory} model, a Helmholtz decomposition was used to write the internal velocity field as $u(x,y,t)=\phi_{x}-A_{y}$ and $v(x,y,t)=\phi_{y}+A_{x}$, where $\phi$ is the velocity potential and the function $A$ contributes to the vorticity field $\omega=v_{x}-u_{y}=A_{xx}+A_{yy}$. For our formulation of travelling solutions that are steady in a co-moving frame, $A(x,y)$ satisfies the following boundary-value problem,
\begin{subequations}\label{eq:1}
\begin{align}
\label{eq:1a}
-\frac{\partial A}{\partial x} = \frac{1}{Re}\bigg( \frac{\partial^2 A}{\partial x^2}+\frac{\partial^2 A}{\partial y^2}\bigg) \qquad &\text{for}~ y\leq \zeta(x), \\
\label{eq:1b}
\frac{\partial A}{\partial x}=\frac{2}{Re}\frac{\partial^2 \zeta}{\partial x^2} \qquad &\text{at}~y=\zeta(x),\\
\label{eq:1c}
A_{x} \to 0, ~A_{y} \to 0 \qquad &\text{as}~y \to -\infty,
\end{align}
\end{subequations}
where $y=\zeta(x)$ is assumed known. In conformal variables, system \eqref{eq:1} becomes
\begin{subequations}\label{eq:2}
\begin{align}
\label{eq:2a}
y_{\xi}A_{\eta}-y_{\eta}A_{\xi}=\frac{1}{Re}(A_{\xi \xi}+A_{\eta \eta}) \qquad &\text{for}~ \eta \leq 0, \\
\label{eq:2b}
X_{\xi}A_{\xi}-Y_{\xi}A_{\eta}=\frac{2}{Re}\frac{(X_{\xi}Y_{\xi \xi}-Y_{\xi}X_{\xi \xi})(X_{\xi}^2+Y_{\xi}^2)}{X_{\xi}^3} \qquad &\text{at}~\eta=0,\\
\label{eq:2c}
A_{\xi} \to 0, ~A_{\eta} \to 0 \qquad &\text{as}~\eta \to -\infty,
\end{align}
\end{subequations}
and the vorticity field is given by $\omega=(A_{\xi \xi}+A_{\eta \eta})/(x_{\xi}^2+y_{\xi}^2)$.
The functions $X(\xi)$ and $Y(\xi)$, in boundary condition \eqref{eq:2b}, parameterise the free surface, $y=\zeta(x)$, and are assumed known from solving system \eqref{eq:steadyequations}. The functions $x(\xi,\eta)$ and $y(\xi,\eta)$, in PDE \eqref{eq:2a} and the expression for $\omega$, are the analytic continuation of these within the flow domain via Laplace's equation, and may be calculated by the Poisson integral formula (or in Fourier space with the multiplier $\mathrm{e}^{2 \lvert k \rvert \pi \eta}$). As an example, if we know the coefficients of the Fourier series $Y(\xi)=\sum_{k=-\infty}^{\infty} c_k \mathrm{e}^{2 k \pi \mathrm{i}\xi}$, we may then calculate $y(\xi,\eta)=\eta+\sum_{k=-\infty}^{\infty} c_k \mathrm{e}^{2 \lvert k \rvert \pi \eta} \mathrm{e}^{2 k \pi \mathrm{i}\xi}$.

While both of systems \eqref{eq:1} and \eqref{eq:2} are linear, they are difficult to solve: \eqref{eq:1} on account of evaluation of boundary condition \eqref{eq:1b} at the free surface $y=\zeta$, and \eqref{eq:2} due to the coefficients being known nonlinear functions.
We now proceed to solve for the vorticity field analytically, first by constructing linear solutions for the free surface in \S\,\ref{sec:boringlinear}, and then for the function $A$ and the vorticity field in \S\,\ref{sec:boringlinear2}.

\subsubsection{Linear theory for the free surface}\label{sec:boringlinear}
We now analytically study small amplitude solutions for the unknown free surface. These will be required for our small amplitude study of the vorticity field, and will also provide a possible explanation for the behaviour of solution branches that we later observe numerically in the nonlinear regime in \S\,\ref{sec:steadyresults}.

We consider a Stokes expansion in powers of a small amplitude parameter, $\ep$. In substituting for solutions of the form $X \sim \xi + \ep X_1$, $Y \sim \ep Y_1$, $\Psi \sim \ep \Psi_1$, and $\Phi\sim \ep \Phi_1$ into equations \eqref{eq:steadyequationsa} and \eqref{eq:steadyequationsb}, at $O(\ep)$ two equations are found for the first-order perturbation solutions. We eliminate $\Phi_1$ from these two equations by using the harmonic relation $\Phi_1=-\mathscr{H}[\Psi_1]$, and then eliminate $\Psi_1$ to find the single equation
\begin{equation}\label{eq:linearsingleeq}
X_{1 \xi} - \frac{Y_1}{F^2}-\frac{P}{F^2}Y_{1 \xi} +\frac{B}{F^2}Y_{1 \xi \xi} +\frac{4}{Re}X_{1 \xi \xi} + \frac{4}{Re^2}X_{1 \xi \xi \xi}=0.
\end{equation}

In writing $Y(\xi)$ as a Fourier series of the form
\begin{equation}\label{eq:YFourierseries}
Y(\xi)=a_0+\sum_{k=1}^{\infty} [a_k\cos{(2k \pi \xi)} + b_k \sin{(2k \pi \xi)}],
\end{equation}
we use the harmonic relation $X_1=-\mathscr{H}[Y_1]$ and equate each coefficient of $\cos(2 k \pi \xi)$ and $\sin{(2 k \pi \xi)}$ to zero to find
\begin{equation}\left. \quad
\begin{aligned}\label{eq:fouriercoefficients}
 \bigg(2k \pi -\frac{1}{F^2} -\frac{4 k^2 \pi^2 B}{F^2}-\frac{32 k^3\pi^3}{Re^2}\bigg)a_k + \bigg(\frac{16 k^2 \pi^2}{Re}-\frac{2 k \pi P }{F^2}\bigg)b_k=0,\\
 \bigg(2k \pi -\frac{1}{F^2} -\frac{4 k^2 \pi^2 B}{F^2}-\frac{32 k^3\pi^3}{Re^2}\bigg)b_k - \bigg(\frac{16 k^2 \pi^2}{Re}-\frac{2 k \pi P}{F^2} \bigg)a_k=0.
\end{aligned}\quad \right\}
\end{equation}

For non degenerate solutions, there must exist a value of $k$ for which at least one of $a_k$ or $b_k$ is non zero. 
This yields
\begin{subequations}\label{eq:lineardisprel}
\refstepcounter{equation}\label{eq:lineardisprelA}
\refstepcounter{equation}\label{eq:lineardisprelB}
\begin{equation}
 2k \pi -\frac{1}{F^2} -\frac{4 k^2 \pi^2 B}{F^2}-\frac{32 k^3\pi^3}{Re^2}=0 \qquad \text{and} \qquad
P=\frac{8 k \pi F^2}{Re}.
\tag{\ref*{eq:lineardisprelA},$b$}
\end{equation}
\end{subequations}
In the inviscid regime, for which $1/Re$ and $P$ are both zero, it is possible to find values of $B$ and $F$ for which \eqref{eq:lineardisprelA} is satisfied for two values of $k$. These are known as Wilton's ripples, after \cite{wilton_1915}. However, when $1/Re$ and $P$ are non-zero this is not possible, due to condition \eqref{eq:lineardisprelB}.

This seems to suggest that the discrete solution branch structure discovered in the inviscid problem, shown in figure~\ref{fig:sheltonetal}$(a)$, will not occur in our current viscous formulation. This is because in the beyond-all-orders asymptotic theory of \cite{shelton2022exponential}, the values of the Bond number dividing adjacent solution branches (shown with black circles in figure~\ref{fig:sheltonetal}$(a)$) were determined from the failure of a solvability condition. As the linear regime was approached ($\mathscr{E} \to 0$), these values of the Bond number tended towards those determined by \cite{wilton_1915}. Thus, the lack of these linear Wilton ripples in our current viscous formulation suggests that the discrete branch structure of nonlinear solutions may not persist in the small surface tension limit, which was the asymptotic limit under which \cite{shelton2022exponential} obtained the solvability condition.

\subsubsection{Linear theory for the vorticity field}\label{sec:boringlinear2}
We now consider linear solutions to the conformal formulation for $A(\xi,\eta)$ and the vorticity field $\omega=(A_{\xi \xi}+A_{\eta \eta})/(x_{\xi}^2+y_{\xi}^2)$.
We substitute $x\sim \xi + \ep x_1$, $y \sim \eta + \ep y_1$, and $A \sim \ep A_1$, into the conformal system \eqref{eq:2}.
At $O(\ep)$, this yields the PDE $A_{1\xi}=-(A_{1 \xi \xi}+A_{1 \eta \eta})/Re$ for $\eta \leq 0$, boundary condition $A_{1 \xi}=2 Y_{1 \xi \xi} /Re$ at $\eta=0$, and decay conditions $A_{1 \xi} \to 0$, $A_{1 \eta} \to 0$ as $\eta \to -\infty$. The linear vorticity field may then subsequently be determined as $\omega \sim \ep (A_{1 \xi \xi}+A_{1 \eta \eta})$.
We solve the boundary-value problem for $A_1$ by separation of variables, which yields a Fourier series expansion in $\xi$, with coefficients depending on $\eta$. Two of the four coefficients must be zero to satisfy the decay conditions as $\eta \to -\infty$, and evaluation of the boundary condition at $\eta=0$ relates the remaining coefficients to those for the free surface Fourier expansion from \eqref{eq:YFourierseries}.
Overall, this yields our solution for the vorticity field as
\begin{equation}
\begin{aligned}
\omega \sim \ep \sum_{k=1}^{\infty} (2k\pi)^2\Big[a_k \Big{\{ } \mathrm{e}^{\sqrt{2k\pi(2k\pi+\i Re)}\eta}+c.c.\Big{\} }+ \i b_k \Big{ \{ }\mathrm{e}^{\sqrt{2k\pi(2k\pi+\i Re)}\eta}-c.c. \Big{ \} }\Big] \cos{(2 k \pi \xi)}\\
-(2k\pi)^2\Big[\i a_k \Big{\{ } \mathrm{e}^{\sqrt{2k\pi(2k\pi+\i Re)}\eta}-c.c.\Big{\} } -b_k \Big{\{ } \mathrm{e}^{\sqrt{2k\pi(2k\pi+\i Re)}\eta}+c.c.\Big{\} }\Big]\sin{(2 k \pi \xi)},
\end{aligned}
\end{equation}
which decays with the behaviour $\exp \big [ \big( k \pi +[(2k\pi)^2+Re^2]^{1/2}/2\big)^{1/2}\eta \big]$ for $\eta \leq 0$.

\section{Steadily travelling solutions}\label{sec:steadyresults}
In this section, we numerically calculate steadily solutions of our viscous formulation \eqref{eq:steadyequations} for small values of the surface tension, $B$. These correspond to travelling wave solutions that are steady in a comoving frame.

\subsection{The steady numerical method}\label{sec:numericalsteady}
We will use an iterative method to solve system \eqref{eq:steadyequations} with Newton's method, in which each equation is evaluated at collocation points in the conformal variable, $\xi$. Each derivative and Hilbert transform are efficiently evaluated in physical space by first utilising spectral relations for the Fourier multiplier of each operator in Fourier space.

We begin by assuming that an initial guess for $Y(\xi)$ is known at each of the $N$ collocation points $\xi_{l}=-1/2+l/N$. In practice, this is either a linear solution from \S\,\ref{sec:boringlinear} or, for more energetic solutions, a previously computed numerical solution with different parameter values. When combined with the unknown constants, the Froude number $F$ and nondimensional wind strength $P$, we have a total of $N+2$ unknown constants. 

Each component of the governing equations is then evaluated efficiently by using spectral relations in Fourier space combined with the fast Fourier transform algorithm. Since the Fourier symbol for differentiation is $2 \pi \mathrm{i} k$, and that for the Hilbert transform is $\mathrm{i} \cdot \text{sgn}(k)$, we have that $Y_{\xi}=\mathscr{F}^{-1}[2 \pi \mathrm{i} k \mathscr{F}[Y]]$, and $\mathscr{H}[Y]=\mathscr{F}^{-1}[\mathrm{i} \cdot \text{sgn}(k) \mathscr{F}[Y]]$. Here, $\mathscr{F}$ denotes the Fourier transform. Since an initial guess for $Y$ is known, we first calculate $X_{\xi}$ from the harmonic relation \eqref{eq:steadyharmonicrelationsC}, then $\Psi_{\xi}$ from equation \eqref{eq:steadyequationsb}, and lastly $\Phi_{\xi}$ from \eqref{eq:steadyharmonicrelationsD}. 

Evaluation of our amplitude condition, the energy \eqref{eq:energy}, requires knowledge of $\Phi$. This is determined spectrally with the Fourier multiplier of integration, $1/(2 \pi \mathrm{i} k)$ for $k \geq 1$ and $0$ if $k=0$. Condition \eqref{eq:phicond}, which correctly determines the constant of integration of $\Phi$, is enforced in Fourier space by setting the constant level of $\Phi X_{\xi}$ to zero.

We then evaluate Bernoulli's equation \eqref{eq:steadyequationsa} at the $N$ collocation points $\xi_{l}=-1/2+l/N$, and the energetic constraint \eqref{eq:energy}, which yields $N+1$ conditions, for the $N+2$ unknowns. We note that since solutions to the current formulation are translationally invariant, a further condition could be imposed to remove this and close the system. However, we have found that numerical convergence is faster when solving the underdetermined system with the Levenberg-Marquit algorithm.
We then seek to minimise the square of the $l^2$ norm such that it is smaller than $10^{-10}$.

For the numerical solutions presented in the following sections, we have used $N=512$ in \S\,\ref{sec:fixedvisc} where $Re$ is fixed, and $N=1024$ in \S\,\ref{sec:distvisc} where $Re$ depends on $B$. Solutions were calculated on a desktop computer using fsolve in MATLAB, which usually took under a second to converge with approximately $10$ iterations. The residual was also typically minimised below $10^{-13}$. However for some more difficult solutions, such as those near the end points of branches (solutions $a$ and $b$ in figure~\ref{fig:bifurc} for instance), up to $100$ iterations were required for the residual to fall below $10^{-10}$.

\begin{figure}
\centering
\includegraphics[scale=1]{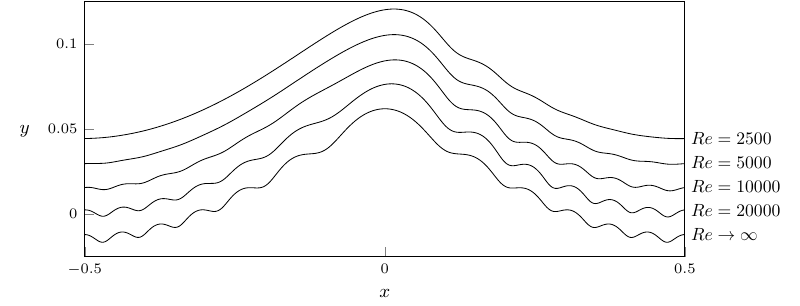}
\caption{\label{fig:solutions}  The effect of increasing viscosity is shown, starting from an inviscid solution found as $Re \to \infty$, with $B=0.002278$, $F=0.4307$, and $\mathscr{E}=0.4$. As the viscosity increases, asymmetry develops in the parasitic capillary ripples, which are most noticeable near the forward face of the travelling wave. These solutions have $Re=\{20000,10000,5000,2500\}$, $F=\{0.4307,0.4308,0.4310,0.4310\}$, and $P=\{0.001424,0.001667,0.001770,0.002574\}$. For visibility, each profile has been shifted vertically by $0.015$.}
\end{figure}

\subsection{Viscous solutions with small surface tension}\label{sec:fixedvisc}

To demonstrate the effect that viscosity has on our solution profiles, we first take an inviscid gravity-capillary wave (determined in the current formulation as $Re \to \infty$), and then use this as an initial guess to converge upon a numerical solution with a finite value of $Re$. The results of this are shown in figure~\ref{fig:solutions}, in which we start with a gravity-capillary wave with $B=0.002278$, $F=0.4307$, and $\mathscr{E}=0.4$. We then compute solutions with the same value of $B$ and $\mathscr{E}$, but different values for the Reynolds number, given by $Re=20000$, $Re=10000$, $Re=5000$, and $Re=2500$. We see that the inviscid profile is symmetric about the wave crest, with parasitic capillary ripples distributed across both the forward and rear faces of the travelling wave. As the effect of viscosity increases (corresponding to decreasing $Re$), asymmetry is seen to develop, and the capillary ripples become less noticeable on the rear face of the wave. The surface profile of these viscous gravity capillary waves closely resembles the experimental observations by \cite{ebuchi_1987} and \cite{perlin1993parasitic}, although we note that this latter study produced capillary ripples whose amplitude fluctuated in time.

\begin{figure}
\centering
\includegraphics[scale=1]{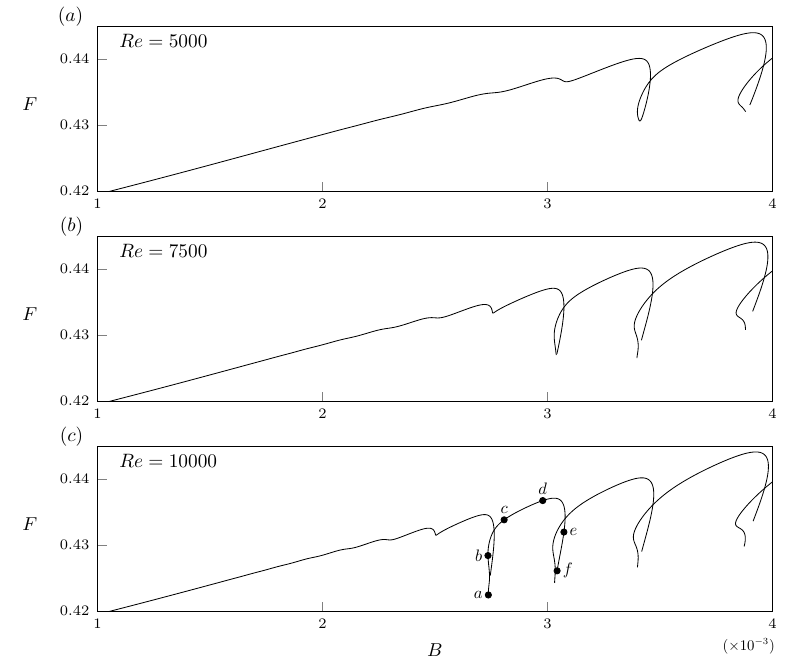}
\caption{\label{fig:bifurc} Branches of solutions are shown in the $(B,F)$ plane for fixed energy, $\mathscr{E}=0.4$. Each subfigure shows the solution branches computed for a different fixed value of the Reynolds number, which are $(a)$ $Re=5000$, $(b)$ $Re=7500$, and $(c)$ $Re=10000$. The labelled points along the branch in $(c)$ are the location of the solutions plotted in figure~\ref{fig:bifurcsolutions}.}
\end{figure}

The bifurcation structure associated with these viscous gravity-capillary waves is shown in figure~\ref{fig:bifurc} for fixed energy and viscosity. We have fixed $\mathscr{E}=0.4$, and chosen the three values of $Re=10000$, $Re=7500$, and $Re=5000$ to explore. Given one solution, the surrounding branch is explored in the $(B,F)$ plane by numerical continuation. We have focused on detecting the branches of solutions that exist for small values of the surface tension parameter, $B$. For inviscid solutions, a detailed bifurcation structure emerges in the small surface tension limit, shown in figure~\ref{fig:sheltonetal}$(a)$, in which a countably infinite number of adjacent solution branches pile up as $B \to 0$. The same phenomena is not seen to occur in figure~\ref{fig:bifurc}, in which the viscosity is held constant as the Bond number decreases. We see in figure~\ref{fig:bifurc}$(c)$ that while these adjacent branches do exist for $B>0.003$, they quickly disappear in the limit of $B \to 0$. Furthermore, as the effect of viscosity increases, the fingering structure of the solution branches disappears at larger values of $B$. When $Re=5000$, the discrete branch structure occurs for $B>0.004$. 
\begin{figure}
\centering
\includegraphics[scale=1]{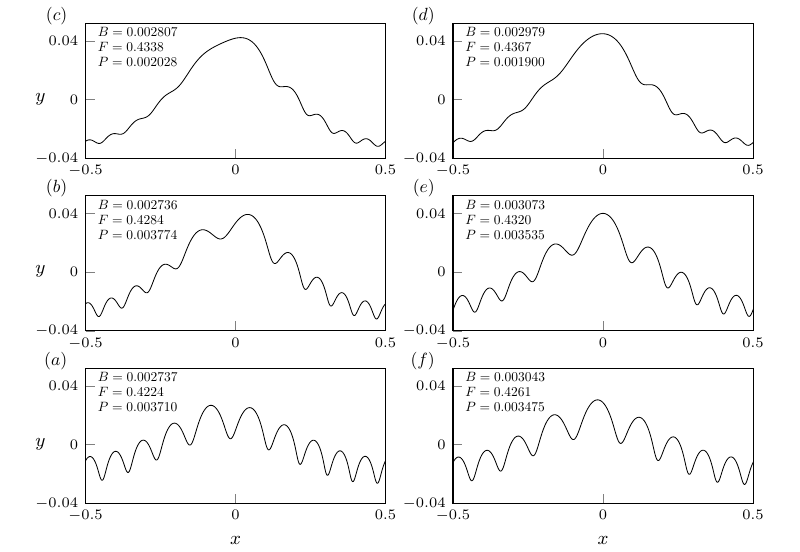}
\caption{\label{fig:bifurcsolutions} The free surface, $y=\zeta(x)$, is shown for six numerical solutions across the same solution branch from figure~\ref{fig:bifurc}$(c)$. These solutions have $\mathscr{E}=0.4$ and $Re=10000$. Solutions $(a)$ and $(f)$ correspond to where each side of the solution branch terminated, beyond which no further solutions could be obtained through numerical continuation.}
\end{figure}
The surface profiles of solutions labeled $a$ to $f$ in figure~\ref{fig:bifurc}$(c)$, with $Re=10000$, are shown in figure~\ref{fig:bifurcsolutions}. The solutions shown in $(c)$ and $(d)$, which lie at the top of the solution branch in figure~\ref{fig:bifurc}$(c)$, resemble the parasitic ripples observed to form physically on steep travelling Stokes waves. They also have smaller values of the wind strength, $P$, than the other displayed solutions, and thus are expected to be more physically realisable.

\subsection{Distinguished limit between viscosity and surface tension}\label{sec:distvisc}
In the previous section, we fixed the Reynolds number, $Re$, and computed solution branches as the Bond number, $B$, was varied. It was seen that the fingering branch structure, associated with inviscid solutions for small surface tension (figure~\ref{fig:sheltonetal}$(a)$), does not emerge when $B$ is decreased and $Re$ held constant. The discrete branching structure, which occurs as $B \to 0$ in the inviscid regime is associated with the exchange of energy between solutions dominated by gravitational energy (at the top of each solution branch), and solutions dominated by capillary energy (down the sides of each solution branch), as measured by each component of \eqref{eq:energy}. In this section, we explore the same bifurcation structure in the $(B,F)$ plane, but with a specified scaling for $Re$, given by
\begin{equation}\label{eq:distRe}
Re=\frac{\lambda_{\alpha}}{B^{\alpha}}.
\end{equation}
The intention of this choice is to investigate possible distinguished limits of $Re$ for which viscosity has the same effect as capillarity on the parasitic capillary ripples, and also investigate whether a discrete branching structure can emerge as $B \to 0$.

We compute the bifurcation structure of solutions, between $B=0.001$ and $0.002$, for $\alpha=1$, $2$, and $3$. The constant $\lambda_{\alpha}$ in relationship \eqref{eq:distRe} is specified such that the distinguished curve passes through the point $B=0.005$ and $Re=5000$, which yields $\lambda_{\alpha}=5000 \times 0.005^{\alpha}$.
\begin{figure}
\centering
\includegraphics[scale=1]{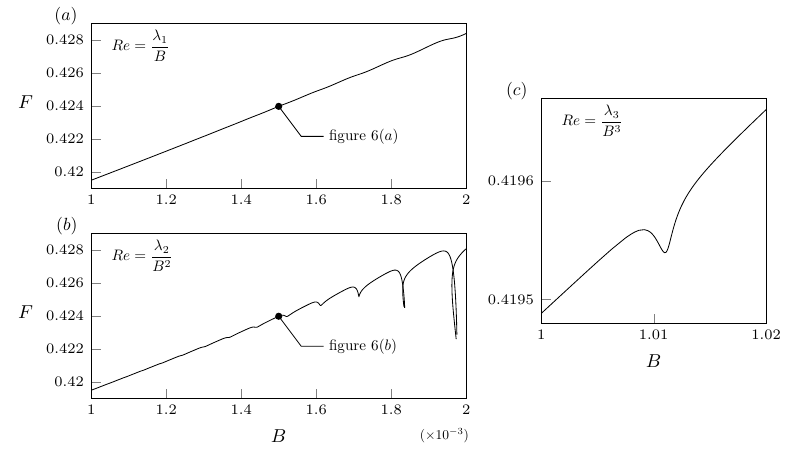}
\caption{\label{fig:distlimit} Branches of solutions are shown in the $(B,F)$ plane for fixed energy, $\mathscr{E}=0.4$. The Reynolds number, $Re$, is specified by $Re=\lambda_{\alpha}/B^{\alpha}$ from \eqref{eq:distRe}, and thus the effect of viscosity decays proportionally to the surface tension. In $(a)$ we have $\alpha=1$, $(b)$ $\alpha=2$, and $(c)$ $\alpha=3$. Note that $\lambda_1=25$, $\lambda_2=0.125$, and $\lambda_3=0.000625$ are chosen such that the distinguished limit \eqref{eq:distRe} intersects with $Re=5000$ and $B=0.005$. The marked points correspond to the solutions shown in figure~\ref{fig:ripples}.}
\end{figure}
These solution branches are shown in the $(B,F)$ bifurcation diagram of figure~\ref{fig:distlimit}, in which subfigure~$(a)$ has $\alpha=1$, $(b)$ has $\alpha=2$, and $(c)$ has $\alpha=3$. The effect of viscosity is largest in the solutions along branch $(a)$. We see that in all three cases, the discrete branching structure recedes as $B$ decreases, such that below a certain value of $B$, all solutions found by numerical continuation are dominated by the effect of gravity. The parasitic capillary ripples present in these solutions appear as a small perturbation to the base nonlinear viscous-gravity wave. Even with $\alpha=3$ in figure~\ref{fig:distlimit}$(c)$, the branch of solutions can be continued to $B=0$ in this manner.

We note that numerical verification of these trends for larger values of $\alpha$ would require computation of the bifurcation diagram to smaller values of $B$ to check if the discrete branching structure persists. However, due to the beyond-all-order (as $B \to 0$) nature of the parasitic capillary ripples, there exists a value of $B$ below which these are not captured by double precision accuracy. When this occurs, the vertical branches of the bifurcation diagram (in which energy is transferred into the oscillatory ripples) are unable to be computed through numerical continuation from the gravity dominated solutions. In practice, for $\mathscr{E}=0.4$ this occurs when $B\leq 0.0008$. It is possible that the discrete branching structure of solutions is only recovered in the small-surface-tension limit when the effect of viscosity, as measured by $1/Re$, is exponentially small in comparison to $B$. However, due to the lack of any multiple-scales asymptotic theory for this regime, even in the inviscid framework, there is no analytical evidence to support this.

\begin{figure}
\centering
\includegraphics[scale=1]{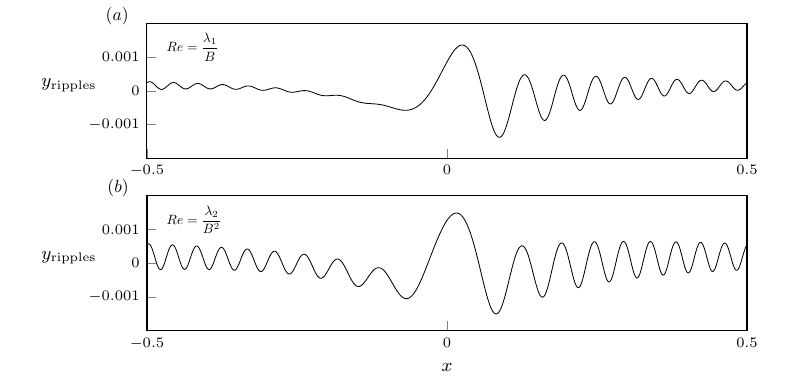}
\caption{\label{fig:ripples} The parasitic capillary ripples present in the free surface are shown for two of the solutions with $B=0.0015$ in the bifurcation diagram of figure~\ref{fig:distlimit}. These profiles have been estimated numerically by calculating $y_{\text{ripples}}=Y-Y_0-BY_1$, where $Y_0(\xi)$ is the solution found with $B=0$, and $Y_1(\xi)$ has been estimated from $(Y-Y_0)/B$ with $B=0.0005$. The effect of viscosity is stronger in solution $(a)$, which produces substantial asymmetry in the parasitic ripple profile.}
\end{figure}
The parasitic capillary ripples present in the solution profile are shown in figure~\ref{fig:ripples} for $\alpha=1$ and $\alpha=2$. These were computed with $B=0.0015$, and correspond to the marked locations in figure~\ref{fig:distlimit}$(a)$ and \ref{fig:distlimit}$(b)$. In this figure, we have subtracted the leading- and first-order solutions, $Y_0(\xi)$, and $BY_1(\xi)$, and to view the ripples in more detail, defined
\begin{equation}\label{eq:rippleestimate}
y_{\text{ripples}} = Y(\xi)-Y_0(\xi)-BY_1(\xi).
\end{equation}
The leading-order solution, $Y_0$, was computed numerically with $B=0$ and $\mathscr{E}=0.4$. For the first-order solution, rather than substituting asymptotic expansions into the governing system \eqref{eq:steadyequations} and then solving the $O(B)$ equation numerically, we have estimated this by computing $Y_1 \sim (Y-Y_0)/B$ with $B=0.0005$. We see in figure~\ref{fig:ripples}$(a)$ that the ripple amplitude is highly asymmetric under this specification of $Re$, and that the ripple wavelength slightly decreases as we travel to the right away from the wavecrest. When the effect of viscosity is weaker, as in \ref{fig:ripples}$(b)$ with $\alpha=2$, the asymmetry is less prominent.

This behaviour is expected when analysing our previous exponential asymptotic theory for nonlinear gravity-capillary waves. This is because the inviscid parasitic ripples are generated by the Stokes phenomenon, which arises from branch points in the analytic continuation of the leading-order gravity wave solution, and this yields their exponential scaling of
\begin{equation}\label{eq:WKB}
y_{\text{ripples}} \sim A(\xi)\mathrm{e}^{-\chi(\xi)/B} +c.c.,
\end{equation}
\begin{figure}
\centering
\includegraphics[scale=1]{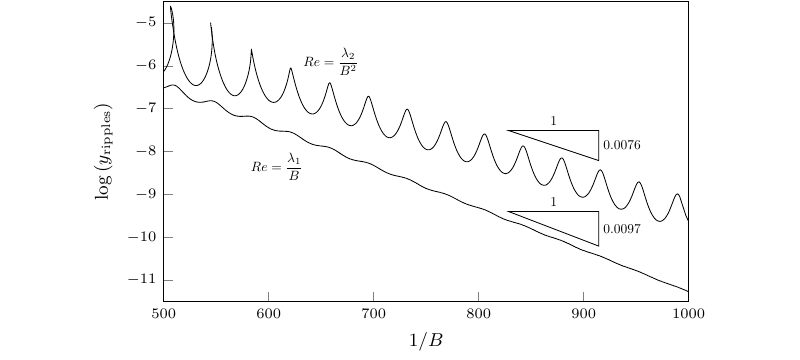}
\caption{\label{fig:expscaling} The parasitic capillary ripple amplitude, measured near $\xi=0.5$, is shown against the Bond number. This amplitude is shown for each solution from the branches in figure~\ref{fig:distlimit}$(a)$ and $(b)$. The linear behaviour in a semi-log plot is indicative that this amplitude is exponentially-small as $B \to 0$. For $Re=\lambda_1/B$ in $(a)$, the gradient is approximately $-0.0097$, and for $Re=\lambda_2/B^2$ it is $-0.0076$. This is compared with the inviscid asymptotic theory of \cite{shelton2022exponential}, which yields a gradient of $-0.0077$.}
\end{figure}
where $c.c.$ denotes complex conjugate. Here, $A$ is an amplitude function, and $\chi$ controls the exponential scaling as $B \to 0$. When $Re=O(1/B)$ as in figure~\ref{fig:ripples}$(a)$, the viscous terms in the governing equations \eqref{eq:steadyequations} will feed into the differential equations for both $A$ and $\chi$. However, when $Re=O(1/B^2)$ as in \ref{fig:ripples}$(b)$, the viscous terms will only alter the amplitude equation for $A$. 

This is demonstrated in the semi-log plot of figure~\ref{fig:expscaling}, in which the parasitic ripple amplitude is shown against $1/B$, for $\alpha=1$ and $\alpha=2$. The slope of this behaviour corresponds to the exponential scaling in \eqref{eq:WKB} evaluated at $\xi=1/2$, near to which the amplitude was measured, which is given by $-\chi(1/2)$. For inviscid capillary ripples, this gradient was determined analytically as $-0.0077$, as shown in figure~\ref{fig:sheltonetal}$(c)$, which is very close to the measurement of $-0.0076$ for $Re=O(1/B^2)$ in figure~\ref{fig:expscaling}. However, when $Re=O(1/B)$, this gradient is steeper with a value of $-0.0097$. We also note that for $Re=O(1/B^2)$ in figure~\ref{fig:expscaling}, all solutions along the branch have this asymptotic scaling as $B \to 0$, in contrast to that in the inviscid regime (see the spikes in figure~\ref{fig:sheltonetal}$(c)$). It is therefore expected that the beyond-all-order solvability mechanism dividing adjacent branches of inviscid solutions \citep{shelton2022exponential} will not manifest in the future exponential asymptotic analysis of our current viscous formulation.

\section{Time-dependent results}\label{sec:timeresults}

We now study the temporal stability of the steady solutions found in \S\,{\ref{sec:steadyresults}}.
Firstly, in \S\,\ref{sec:linearstability} we analyse the linear stability of these solutions. Growth rates are obtained for small superharmonic perturbations to the steady solutions by solving a linear eigenvalue problem for the perturbations. The eigenvalue, $\sigma$, corresponds to the growth rate of the perturbations which contain a term of the form $\mathrm{e}^{\sigma t}$. We will find that the steady solutions are super-harmonically stable since $\text{Re}[\sigma] \leq 0$.
Secondly, in \S\,\ref{sec:nonlinearstability} we study the nonlinear stability of the solutions.
This is achieved by considering the unsteady time-evolution problem, which was formulated in \S\,\ref{sec:confmaptime} as a set of time-evolution equations for $Y(\xi,t)$ and $\Phi(\xi,t)$ in terms of the conformal variable $\xi$.
We will pick an initial condition at $t=0$, and for $t>0$ set all associated constants, $B$, $F$, $Re$, and $P$, to that of the steady solution whose nonlinear stability is to be studied.
This allows us to comment on the `global' stability and attractiveness of these solutions, as $t \to \infty$, with respect to different initial conditions, such as a steep gravity wave or an almost flat free surface.

\subsection{Linear stability of the steady solutions}\label{sec:linearstability}

We now examine the linear stability of the steady solutions from \S\,\ref{sec:steadyresults}. 
We begin by detailing the methodology used to analyse the linear stability of these steady solutions.
The method presented here is similar to that used by \cite{tiron2012linear} for pure capillary waves and \cite{blyth2022stability} for constant vorticity waves, with the exception that we only study superharmonic modes that fit with within the assumed period $-1/2 \leq \xi \leq 1/2$, in order to draw comparisons with the time-evolution problem studied in \S\,\ref{sec:nonlinearstability}.

\subsubsection{Numerical implementation}\label{sec:stabnumimple}
We begin by linearising about a steady solution by writing $X \sim X_0(\xi) + \ep X_1(\xi,t)$, $Y \sim Y_0(\xi) + \ep Y_1(\xi,t)$, $\Phi \sim \Phi_0(\xi) + \ep \Phi_1(\xi,t)$, and $\Psi \sim \Psi_0(\xi) + \ep \Psi_1(\xi,t)$. Here, $X_0$ $Y_0$, $\Psi_0$, and $\Phi_0$ are solutions of equations (\hyperref[eq:steadyequations]{\ref{eq:steadyequations}$a$--$d$}), whose temporal stability we will analyse by studying growth/decay rates of the perturbation quantities. Substitution of these expressions into the time-dependent equations (\hyperref[eq:Tmapevolveboth]{\ref{eq:Tmapevolveboth}$a$--$d$}) yields at $O(\ep)$ four equations for $X_1$, $Y_1$, $\Phi_1$, and $\Psi_1$. Two of these equations are given by $X_{1 \xi}=-\mathscr{H}[Y_{1 \xi}]$ and $\Psi_{1}=\mathscr{H}[\Phi_{1}]$. The other two equations, obtained at $O(\ep)$ from \eqref{eq:Tmapyevolve} and \eqref{eq:Tmapphievolve}, have coefficients involving the leading-order quantities, which in practise we calculate expressions for with a symbolic programming language. 
Next, for the time-dependent perturbations we pose a Floquet ansatz of the form
\begin{equation}\label{eq:floquetansatz}
\{ X_1,Y_1,\Phi_1,\Psi_1 \}=\mathrm{e}^{\sigma t} \sum_{m=-M}^{M} \{a_m,b_m,c_m,d_m \} \mathrm{e}^{2 m \pi \mathrm{i} \xi},
\end{equation}
where the real part of the complex-valued constant $\sigma$ will correspond to the growth/decay rate in time of the perturbation.
Note that the ansatz \eqref{eq:floquetansatz} would differ if used to analyse the stability of time-dependent solutions, such as time-periodic standing waves whose stability was studied by \cite{wilkening2020harmonic}.
Since $\mathscr{H}[\mathrm{e}^{2 m \pi \mathrm{i} \xi}]=\mathrm{i} \cdot \text{sgn}{(m)}\mathrm{e}^{2 m \pi \mathrm{i} \xi}$, where $\text{sgn}$ is the signum function, we immediately have from the $O(\ep)$ harmonic relations that $a_m=-\mathrm{i}\cdot \text{sgn}(m)b_m$ and $d_m=\mathrm{i}\cdot \text{sgn}(m)d_m$. We note that given a solution $\{\sigma,b_{-M},\ldots,b_{M},c_{-M},\ldots,c_{M}\}$ to the $O(\ep)$ equations, there exists another solution given by $\{ \sigma^{*},b_{M}^{*},\ldots,b_{-M}^{*},c_{M}^{*},\ldots,c_{-M}^{*} \}$, which when combined yield a real-valued solution to the $O(\ep)$ equations. Here, $^{*}$ denotes the complex conjugate.

To solve for the unknowns $\{\sigma,b_{-M},\ldots,b_{M},c_{-M},\ldots,c_{M}\}$, we collocate the two remaining $O(\ep)$ equations at the $N:=2M+1$ points $\xi=-1/2+l/N$, for $l=0,\ldots,N-1$. This yields $2N$ algebraic equations for $2N+1$ unknowns, which we write as the linear system $\sigma \mathbf{L} \mathbf{v} = \mathbf{R} \mathbf{v}$. Here, the eigenvalue $\sigma$ is the temporal coefficient from \eqref{eq:floquetansatz}, $\mathbf{v}=[b_{-M},\ldots,b_{M},c_{-M},\ldots,c_{M}]^{T}$ is a $2N \times 1$ eigenvector consisting of the unknown Fourier coefficients, and $\mathbf{L}$ and $\mathbf{R}$ are $2N \times 2N$ matrices.
The matrix $\mathbf{L}$ contains entries associated with the time-derivative in the $O(\ep)$ equations, given by $\mathbf{L}_{ij}=\mathrm{e}^{2 (j-1-M) \pi \mathrm{i} \xi_i}$ for $1 \leq i \leq N$ and $1 \leq j \leq N$,
 $\mathbf{L}_{ij}=\mathrm{e}^{2 (j-N-1-M) \pi \mathrm{i} \xi_{i-N}}$ for $N+1 \leq i \leq 2N$ and $N+1 \leq j \leq 2N$, and $\mathbf{L}_{ij}=0$ otherwise.
 For details on the construction of $\mathbf{R}$, we refer the reader to either the work of \cite{blyth2016stability} (c.f. their eq. 4.17) for a similar boundary-integral problem, or the MATLAB code in our supplementary material which constructs $\mathbf{R}$ column by column.
 
We solve the generalised eigenvalue problem $\sigma \mathbf{L} \mathbf{v} = \mathbf{R} \mathbf{v}$ in MATLAB with the QZ algorithm. The code that implements this method, and produces the results shown next in \S\,\ref{sec:linearstabresults}, is provided alongside this work as supplementary material. When $M=127$, solving this generalised eigenvalue problem on a standard desktop computer takes approximately $3.5$ seconds.

\subsubsection{Linear stability results}\label{sec:linearstabresults}
We now calculate the growth rates, $\sigma$, and the corresponding eigenvectors, $Y_1(\xi,0)$ for instance, for perturbations to steady solutions previously calculated in \S\,\ref{sec:steadyresults}. The parameter values for the steady solutions, whose linear and nonlinear stability we will investigate in detail, are provided in table~\ref{tab:values}.
\begin{table}
  \begin{center}
\def~{\hphantom{0}}
  \begin{tabular}{cccc}
    $Re$ &   $B$  & $F$   &   $P$ \\[3pt]
         ~5000 &   0.0026  & 0.433693732256569 & 0.002241721973881\\
         ~7500 &  0.0026   & 0.433421267153132 &  0.001841654933803\\
         10000 &  0.0026   & 0.433231130404047 &  0.001522514473361\\
  \end{tabular}
  \caption{Parameter values used in the stability results of figure~\ref{fig:eigenvalues} and the time-dependent simulations of figures~\ref{fig:gravityconvergence} and \ref{fig:flatconvergence}. These were obtained from the steady solutions of \S\,{\ref{sec:steadyresults}}, iterated upon such that the residual, defined by the square of the $L^{2}$-norm of equations \eqref{eq:steadyequations}, is smaller than $10^{-20}$.}
  \label{tab:values}
  \end{center}
\end{table}
We will investigate the effect of viscosity and surface wind forcing on the solution stability. 
Each of these solutions has the same Bond number, $B=0.0026$, and different values of the Reynolds number, which are taken to be $Re=5000$, $Re=7500$, and $Re=10000$. Recall that the magnitude of the surface wind forcing, $P$, was an unknown in the steady formulation. The solution with the largest viscous dissipation, $Re=5000$, has a larger value of $P$ to compensate.

\begin{figure}
\centering
\includegraphics[scale=1]{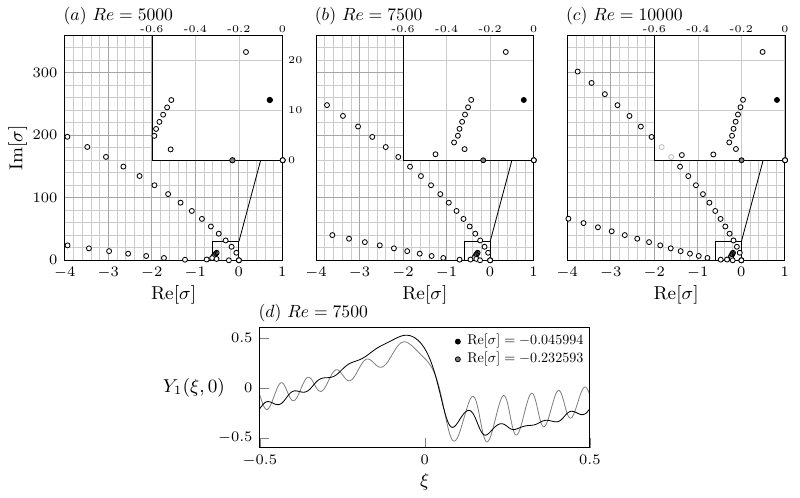}
\caption{\label{fig:eigenvalues} $(a$--$c)$ The complex-valued eigenvalues, $\sigma$, for small amplitude perturbations to the steady solutions from \S\,\ref{sec:steadyresults} are shown. The parameter values for these steady solutions are given in table~\ref{tab:values}, and these eigenvalues were calculated with the numerical methodology described in \S\,\ref{sec:stabnumimple}. From \eqref{eq:floquetansatz}, the real part of $\sigma$ is the temporal growth rate of the perturbation. The eigenvalue shown in black has the largest real part with $\text{Re}[\sigma]= \{ -0.059114 ,-0.045994, -0.0371606 \}$. The first eigenvalue on the real axis, with $\text{Im}[\sigma]=0$, is shown in grey with values $\text{Re}[\sigma]= \{ -0.230998,-0.232593,-0.199846 \}$. These values predict the growth rates observed later in the time-evolution simulations in figures~\ref{fig:gravityconvergence} and \ref{fig:flatconvergence}. $(d)$ Perturbations to the free-surface elevation predicted by the linear stability analysis are shown, which correspond to the eigenvectors for the two labeled eigenvalues in inset $(b)$.}
\end{figure}
The real and imaginary parts of the growth rates, $\sigma$, for perturbations to each of these steady solutions are shown in figure~\ref{fig:eigenvalues}.
While only eigenvalues with $\text{Im}[\sigma] \geq 0$ have been shown in figure~\ref{fig:eigenvalues}, complex-conjugate values are also solutions of the generalized eigenvalue problem.
The real part of these eigenvalues corresponds to the growth rate of time-dependent perturbation.
Note that our formulation does not possess time-reversal symmetry, and therefore there is no four-fold symmetry in the eigenvalue spectrum (in which $\sigma$, $\sigma^*$, $-\sigma$, $-\sigma^*$ would be solutions). Such four-fold symmetry is associated with inviscid formulations that are Hamiltonian \citep{deconinck2011instability}.
Aside from $\sigma=0$, corresponding to translational invariance of the steady formulation (a small perturbation of a steady solution can result in another steady solution), all eigenvalues have negative real part, $\text{Re}[\sigma]<0$.
The solutions are therefore superharmonically stable.

\begin{figure}
\centering
\includegraphics[scale=1]{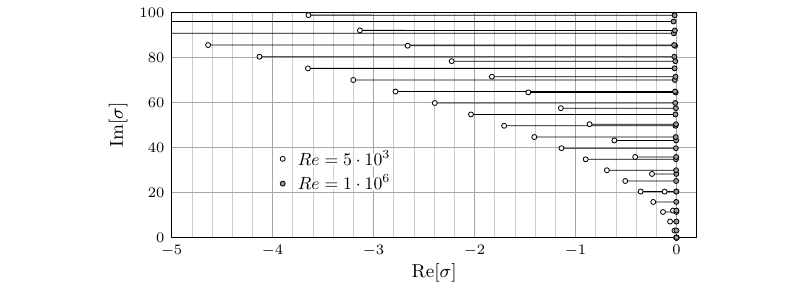}
\caption{\label{fig:eigenvaluesRe} The real and imaginary components of the growth rates, $\sigma$ from \eqref{eq:floquetansatz}, for steady solutions with zero surface tension ($B=0$) are shown. The growth rates shown with white circles have $Re=5 \cdot 10^{3}$, $P=8.229 \cdot 10^{-4}$, and $F=0.4110$. The growth rates shown with grey circles have $Re=1 \cdot 10^{6}$, $P=4.115 \cdot 10^{-6}$, and $F=0.4110$. Black contours between these show the behaviour of the eigenvalues for $50$ intermediary values of $Re$. These tend towards the imaginary axis as $Re \to \infty$.}
\end{figure}

Note that under the inviscid limit of $Re \to \infty$, the growth rates are observed to tend towards the imaginary axis. This is demonstrated in figure~\ref{fig:eigenvaluesRe} for solutions without surface tension. Modulational and high frquency instabilities \citep{creedon2022high} do not emerge here as we only study superharmonic perturbations with zero Floquet exponent in ansatz \eqref{eq:floquetansatz}.

Typically, the eigenvalues with the largest real part will dominate the evolution of time-dependent simulations for solutions close to the steady solution. We investigate this in detail later in \S\,\ref{sec:nonlinearstability}, where it is seen that different choices of initial conditions can result in different rates of convergence for the transient behaviour towards the stable steady state. The growth rates that dominate the time-dependent simulations of \S\,\ref{sec:nonlinearstability} are: (i) the eigenvalue with the largest real part, shown with a black circle in figure~\ref{fig:eigenvalues}, and (ii) the eigenvalue on the real-axis with $\text{Im}[\sigma]=0$, shown with a grey circle in figure~\ref{fig:eigenvalues}.
The steady solution with $Re=7500$, for which the growth rates are shown in figure~\ref{fig:eigenvalues}$(b)$, has $\text{Re}[\sigma]=-0.045994$ and $\text{Re}[\sigma]=-0.232593$ for these two values.

The perturbation solution, $Y_1$, corresponding to each of these two growth rates is shown in figure~\ref{fig:eigenvalues}$(d)$. Here, we have computed the contribution to the free surface elevation, $Y_1(\xi,0)$, from the corresponding eigenvector and the complex-conjugate eigenvector by the relation $Y_1(\xi,0)=\sum_{m=-M}^{M}[b_m \mathrm{e}^{2 m \pi \mathrm{i} \xi}+b_m^{*}\mathrm{e}^{-2m \pi \mathrm{i}\xi}]$ from \eqref{eq:floquetansatz}.
The eigenvector for the $\text{Re}[\sigma]=-0.045994$ mode, shown in black, has more of an effect on the wave amplitude, whereas that for the $\text{Re}[\sigma]=-0.232593$ mode, shown in grey, has a larger contribution to the oscillatory capillary ripples. We now turn to the nonlinear time-evolution problem to observe how these stability results emerge in practice.

We have also studied the linear stability of solutions for larger values of $Re$, to see if a critical Reynolds number exists beyond which instability emerges. However, all solutions we have tested in the range $0 \leq Re \leq 10^{6}$ have been linearly stable.

\subsection{Nonlinear stability of the steadily travelling solutions}\label{sec:nonlinearstability}
We now perform time-evolution simulations to test the nonlinear stability of the steady solutions from \S\,\ref{sec:steadyresults}.
To measure convergence towards the steady solution as $t \to \infty$, we will use the norm $\lvert \mathscr{E}(t)-\mathscr{E} \rvert$, which compares the wave energy \eqref{eq:energy}, $\mathscr{E}(t)$, to that of the intended steady solution, $\mathscr{E}$.
The initial condition we use at $t=0$ will be chosen as one of the following:
\begin{enumerate}[label=(\roman*),leftmargin=*, align = left, labelsep=\parindent, topsep=3pt, itemsep=2pt,itemindent=0pt ]
\item an inviscid gravity wave, which is a solution of the steady equations \eqref{eq:steadyequations} solved for with $\mathscr{E}=0.4$, $B=0$, $P=0$, and $1/Re=0$;
\item a small-amplitude cosine profile, $Y(\xi)=10^{-5}\cos{(2 \pi \xi)}$.
\end{enumerate} 
\captionsetup{width=7in,justification=raggedright}
\begin{sidewaysfigure}
\centering
\vspace{5.2in}
\includegraphics[scale=1]{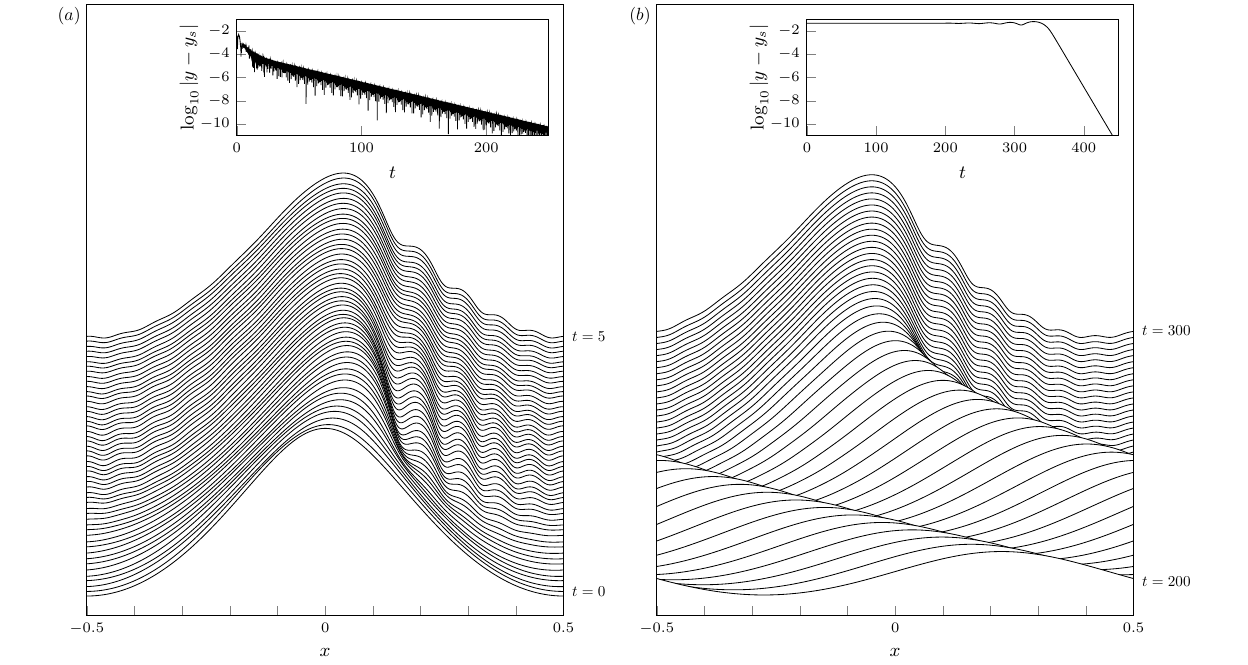}
\caption{\label{fig:timeevolution} The time evolution of the free surface is shown from the initial conditions at $t=0$ of a steep viscous-gravity wave with $\mathscr{E}=0.4$ in $(a)$, and from a linear cosine profile with amplitude $10^{-5}$ in $(b)$. The parameter values used for $t>0$ are given in table~\ref{tab:values} for $Re=5000$, and the time interval between displayed solutions is $0.1$ in $(a)$ and $2$ in $(b)$. The insets show the difference in height of the wavecrest of the steady solution, $y_s$, and that of the current numerical solution, $y$. For the simulation in $(a)$, the solution approaches the steady solution in an oscillatory manner in time, while the solution in $(b)$ approaches the same steady solution monotonically.}
\end{sidewaysfigure}
\captionsetup{width=\textwidth,justification=raggedright}
For $t>0$, we then set the values for $B$, $F$, $Re$, and $P$, to that of a selected steady solution, whose nonlinear stability we wish to analyse. 
The parameter values used in this section are shown in table~\ref{tab:values}.

Surface profiles for both of these time-evolution simulations are shown in figure~\ref{fig:timeevolution}. We see in figure~\ref{fig:timeevolution}$(a)$ that for the initial condition of a steep gravity wave, parasitic ripples form quickly within the interval $0 \leq  t  \leq 5$ due to the nonlinearity of the initial condition. When the initial condition is of smaller amplitude, as in figure~\ref{fig:timeevolution}$(b)$, it takes much longer for the parasitic ripples to form. This is due to the surface wind forcing inducing a small growth rate in the solution, and the ripples then become noticeable within the plotted interval $200 \leq  t \leq 300$. For the simulation with the large amplitude initial condition, the solution tends towards the steady solution in an oscillatory manner, as shown in the inset of figure~\ref{fig:timeevolution}$(a)$, which is associated with a complex-valued growth rate from the linear stability analysis. The simulation with a small amplitude initial condition approaches the steady solution monotonically, which is associated with a real-valued growth rate in the linear stability analysis. These two dominant growth rates were highlighted in figure~\ref{fig:eigenvalues}$(a)$.

\begin{figure}
\centering
\includegraphics[scale=1]{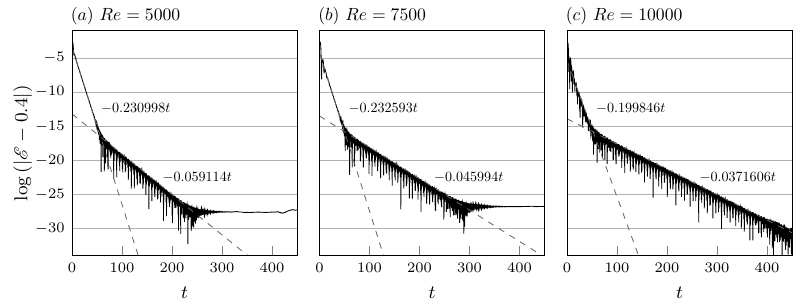}
\caption{\label{fig:gravityconvergence} The magnitude of the difference between the energy, $\mathscr{E}(t)$, and the `target' energy, $0.4$, is shown in a semi-log plot for the three cases of $(a)$ $Re=5000$, $(b)$ $Re=7500$, and $(c)$ $Re=10000$. The initial condition at $t=0$ was chosen to be a steep gravity wave with $\mathscr{E}=0.4$, $B=0$, $P=0$, and $1/Re=0$, and the parameter values used for $t>0$ in each of these simulations are given in table~\ref{tab:values}. The annotated gradients are predictions from the linear stability results of figure~\ref{fig:eigenvalues}.}
\end{figure}
\begin{figure}
\centering
\includegraphics[scale=1]{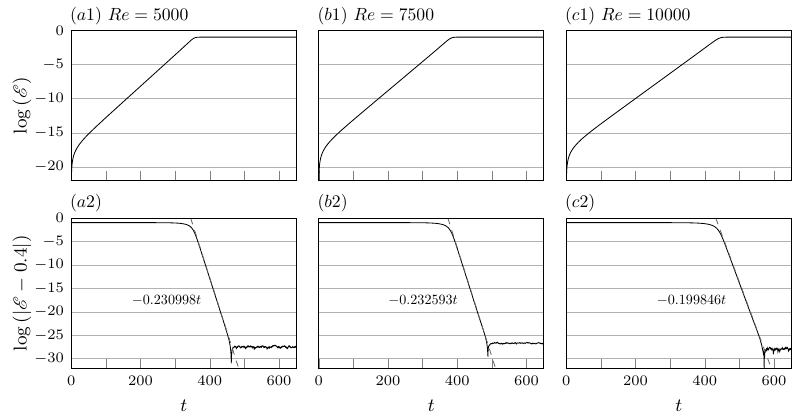}
\caption{\label{fig:flatconvergence} Convergence is shown for the initial condition at $t=0$ of a small amplitude cosine profile, $Y(\xi)=10^{-5}\cos{(2 \pi \xi)}$. The parameter values used in each of these simulations are given in table~\ref{tab:values}, which has $(a)$ $Re=5000$, $(b)$ $Re=7500$, and $(c)$ $Re=10000$. The top row plots $\mathscr{E}(t)$ to show the initial growth rate from the initial condition. The bottom row plots $\lvert \mathscr{E}(t)-0.4 \rvert$ to show the final convergence rate towards the steadily travelling solution. The annotated gradients are predictions from the linear stability results of figure~\ref{fig:eigenvalues}.}
\end{figure}

We begin by examining the stability of three different viscous solutions when starting from an inviscid gravity wave as an initial condition. The difference between the energy of the time-evolved solution, and that of the target solution, is shown in figure~\ref{fig:gravityconvergence}. In each case studied, the energy converges to within the tolerance of the computed steady solution, which occurs at approximately $t=250$ when $Re=5000$, $t=300$ when $Re=7500$, and $t=400$ when $Re=10000$. Convergence is seen to occur quicker when the effect of viscosity, and the associated wind strength $P$, is larger. Note that in each of the three cases shown in figure~\ref{fig:gravityconvergence}, while the initial gravity wave at $t=0$ has $\mathscr{E}=0.4$, this value increases once the associated parameters are set to the values shown in table~\ref{tab:values}. For instance, at the first time-step we have in $(a)$ $\mathscr{E}=0.4442$, $(b)$ $\mathscr{E}=0.4439$, and $(c)$ $\mathscr{E}=0.4437$. Two stages of convergence are observed in figure~\ref{fig:gravityconvergence}. First, there is the fastest stage during which the norm, $\log ( \lvert \mathscr{E}-0.4 \rvert )$, decreases to $-16$, and then there is the slower stage during which the energy oscillates about $0.4$, with the norm decreasing from $-16$ to $-27$. The gradients for each of these stages, predicted by the linear stability analysis of \S\,\ref{sec:linearstability}, are shown within each figure. Both the fast and the slow stages of convergence are predicted by the real part of the eigenvalues, $\sigma$, obtained from this stability analysis.
The fast growth rate in the initial stage of convergence is associated with the largest real-valued eigenvalue ($\text{Im}[\sigma]=0$). For the $Re=7500$ simulation in figure~\ref{fig:gravityconvergence}$(b)$, this is $\text{Re}[\sigma]=-0.232593$.
The slow growth rate in the last stage of convergence is associated with the complex-valued eigenvalue ($\text{Im}[\sigma] \neq 0$) whose real part is largest. For the $Re=7500$ simulation this is $\text{Re}[\sigma]=-0.045994$.
Note that in each of the three simulations shown in figure~\ref{fig:gravityconvergence}, the differing values at which $\mathscr{E}-0.4$ plateaus is due to the accuracy of the parameter values given in table~\ref{tab:values}.

Convergence towards the same steady solutions is also studied in figure~\ref{fig:flatconvergence} from a small-amplitude initial condition at $t=0$. In subfigures $(a1)$, $(b1)$, and $(c1)$, we plot the behaviour of $\log(\mathscr{E})$ in time to show the initial growth rate away from the linear solution. We then plot in $(a2)$, $(b2)$, and $(c2)$ the behaviour of $\log(\lvert \mathscr{E}-0.4 \rvert)$ in time to show the rate of convergence towards the steady solution.
The total time taken for $\log (\lvert \mathscr{E}-0.4\rvert)$ to reach $-27$ is longer than that for the simulations in figure~\ref{fig:gravityconvergence}, which is due to the duration of the initial growth away from the linear profile at $t=0$.
Once $ \lvert \mathscr{E}-0.4 \rvert$ is smaller than $10^{-2}$, the steadily travelling solution is rapidly approached. 
The rate of convergence here is also predicted by the linear stability analysis of \S\,\ref{sec:linearstability}.
However, unlike the transient behaviour in figure~\ref{fig:gravityconvergence} which contained two different growth rates, only one growth rate emerges for the simulations in figure~\ref{fig:flatconvergence}.
This growth rate is not associated with the complex-valued eigenvalue with largest real part (which emerged as a transient in figure~\ref{fig:gravityconvergence}), but rather the largest real-valued eigenvalue.
For instance, the rate of convergence in the simulation with $Re=7500$ in figure~\ref{fig:flatconvergence} is controlled by the real-valued eigenvalue $\sigma=-0.232593$, and no transient behaviour associated with $\text{Re}[\sigma]=-0.045994$ is present.

In summary, the growth rates that emerge in this time-evolution problem can be determined from the formal linear stability analysis of \S\,\ref{sec:linearstability}. However, it is not clear in advance which growth rate will emerge as the dominant transient behaviour, and this depends on the choice of initial condition. The large-amplitude initial condition initially produces fast convergence, which subsequently is dominated by a slower mode (figure~\ref{fig:gravityconvergence}), whereas the convergence that emerges from the small-amplitude initial condition only contains the fast mode (figure~\ref{fig:flatconvergence}).
\begin{figure}
\centering
\includegraphics[scale=1]{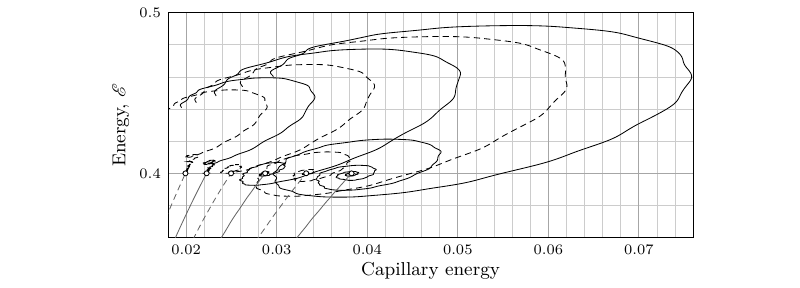}
\caption{\label{fig:phaseenergy} Phase space diagram for time-dependent simulations. We have shown the  energy of the solution, $\mathscr{E}$ from \eqref{eq:energy}, against the capillary energy, defined by the middle term in the integrand on the right-hand side of \eqref{eq:energy}. The white circles indicate the position of six steady solutions with $\mathscr{E}=0.4$ and $Re=7500$. These have $B=\{0.00222,0.0024,0.0026,0.00287,0.0033,0.00365\}$, in order from left to right in the figure. Two simulations have been performed for each set of parameter values associated with these steady solutions. Similar to the simulations from figures~\ref{fig:timeevolution},~\ref{fig:gravityconvergence},~\ref{fig:flatconvergence}, one simulation (black lines/dashes) begins from a large amplitude gravity wave with $\mathscr{E}=0.4$, and the other (grey lines/dashes) from a small-amplitude cosine profile.}
\end{figure}
Further simulations are shown in an (total energy, capillary energy) phase space in figure~\ref{fig:phaseenergy}. The simulations in this figure have different values of the surface tension parameter, $B$. Those that begin from a large amplitude initial condition converge towards the steady solution in an oscillatory manner, as observed for other parameter values in figure~\ref{fig:gravityconvergence}. The radius of these oscillations decays rapidly as $B$ decreases, which is possibly related to the capillary modes of the steady solution being exponentially small as $B \to 0$ (figure~\ref{fig:expscaling}). Conversely, the simulations with a small amplitude initial condition tend towards the steady solution monotonically.

Note while the simulations we show here all result in convergence, indicating nonlinear stability, this is not true in general. For instance, starting with a small amplitude initial condition with multiple Fourier modes (in contrast to $Y(\xi,0)=10^{-5}\cos{(2 \pi \xi)}$ used in figure~\ref{fig:flatconvergence}) results in general unsteady motion persisting at large time. Similar time-evolution simulations have been performed by \cite{hung2009formation} and \cite{murashige2017numerical}, both of whom used a steep gravity wave as an initial condition. However, these works studied the initial formation of capillary ripples, rather than characterise convergence towards steady solutions as $t \to \infty$ by selecting appropriate parameter values. \cite{hung2009formation} observed that the amplitude of the parasitic capillary ripples fluctuated in time, much like that seen in the experiments of \cite{perlin1993parasitic}. We observe similar behaviour in time-evolution simulations when a different value of the wind strength $P$ is used.

\section{Conclusion \& Discussion}\label{sec:disc}

While we have focused on viscous gravity-capillary waves travelling on water of infinite depth, the formulation with finite depth is also of significant interest. This is because viscosity requires a no-slip condition on the lower surface. A nonlinear potential flow model incorporating the no-slip condition at a lower boundary has previously been developed by \cite{dutykh2007viscous}, by first applying a Helmholtz decomposition to the linear Navier-Stokes equations, and then introducing a viscous boundary layer at the substrate. The dominant effect of viscosity in their model, of $O(Re^{-1/2})$, occurs from the lower boundary condition, rather than that of $O(Re^{-1})$ in the surface boundary conditions.
We therefore expect that the discrete branching structure associated with capillarity in figure~\ref{fig:bifurcsolutions} would be further suppressed by viscosity in the presence of a finite depth fluid.

It may also be possible to support steadily travelling surface wave solutions of a viscous fluid, by considering the effect of gravity on a flow down an inclined plane, rather than the surface wind forcing considered in this paper.
The effect of capillarity and viscosity on the parasitic solutions that would emerge in this formulation remains unknown. More generally, it is also of interest to ask whether these viscous boundary layer techniques may be extended to derive potential flow formulations for water waves in the presence of more complicated lower topography and submerged obstacles. Such formulations have been extensively studied in the inviscid regime by \emph{e.g.} \cite{dagan1972two,dias1989open,lustri2012free,ambrose2022numerical}, for which recirculation zones can emerge with the inclusion of viscosity \citep{biswas2004backward}. The analytical insight available in these viscous potential-flow models would provide a distinct advantage to that from the Navier-Stokes equations with a free boundary.

Due to the assumed periodicity of our solutions, the linear and nonlinear temporal stability results in \S\,\ref{sec:timeresults} should be regarded only as evidence for their superharmonic stability with respect to perturbations that lie within the periodic window. Their modulational stability (with respect to subharmonic perturbations wider than the periodic domain) has not been investigated in this work. While nonlinear gravity waves are known to exhibit the modulational instability \citep{longuet1978instabilitiesSub}, there is the possibility that this can be suppressed by the effect of viscosity, which has been studied in a damped nonlinear Schr\"{o}dinger equation by \cite{segur2005stabilizing}.

Previously \cite{longuet_1995} had obtained approximate solutions for the asymmetric parasitic capillary waves that form upon steep gravity waves with viscosity, which built upon an earlier theory by \cite{longuet1963generation}. However, the methodology used there involves a number of modelling assumptions, rather than a systematic asymptotic study of the governing equations. The direct asymptotic analysis of these, in the limit of $B \to 0$, is therefore of interest. In the inviscid study by \cite{shelton2022exponential,shelton2024model}, the capillary ripples observed on steep gravity-capillary waves were determined asymptotically. These were obtained by resolving the Stokes phenomenon generated from branch points in the analytic continuation of the leading-order gravity wave.
We note also that the solutions we have focused upon in this work are steady in a comoving frame, which may not be the precise manner that they occur physically. In experiments performed by \cite{perlin1993parasitic} in a wavetank with a wave maker, parasitic ripples were seen to travel at the same speed as the propagating gravity wave, but the ripples had an amplitude that varied in time. Beyond-all-order asymptotic techniques are well suited to investigate if such time-dependent behaviour can emerge in this formulation for small surface tension, and the application of our viscous formulation to study this is an exiting area of future research. Here, the leading-order gravity wave solution would be time independent, and the exponentially-subdominant order (the parasitic capillary ripples) and associated Stokes lines would be time dependent.
The future extension of this methodology to our viscous formulation is also expected to yield insight into the competing effects of viscosity and surface tension on the observed ripple asymmetry.

\backsection[Acknowledgements]{We thank the anonymous referees for their suggestions that have improved the clarity and scope of our paper, including the suggestion to perform a linear stability analysis to obtain the decay rates of perturbations to our solutions.}
\backsection[Funding]{ JS and PHT acknowledge support by the Engineering and Physical Sciences Research Council (EPSRC grant no. EP/V012479/1). JS was additionally supported by the Engineering and Physical Sciences Research Council (EPSRC grant no. EP/W522491/1)}
\backsection[Declaration of interests]{The authors report no conflict of interest.}

\appendix
\section{The time-dependant conformal mapping}\label{App:timemap}
We now develop the time-dependant mapping that takes the physical fluid domain $-1/2 \leq x \leq 1/2$ and $-\infty < y \leq \zeta(x,t)$ to the lower-half $(\xi,\eta)$ plane. The time-dependent evolution equations that we derive in this section were previously presented in \S\,{\ref{sec:confmaptime}}. The following method is an extension of that presented by \cite{choi1999exact} and \cite{milewski2010dynamics} for inviscid regimes.

We consider both $x=x(\xi,\eta,t)$ and $y=y(\xi,\eta,t)$ to be functions of the conformal domain. The free-surface variables, previously defined in \eqref{eq:Tmapvariables}, are given by $X(\xi,t)=x(\xi,0,t)$, $\Phi(\xi,t)=\phi(x(\xi,0,t),y(\xi,0,t),t)$, $Y(\xi,t)=\zeta(x(\xi,0,t),t)$, and $\Psi(\xi,t) = \psi(x(\xi,0,t),y(\xi,0,t),t)$.
We will derive four coupled integro-differential equations for these. Our aim is to develop expressions for each of the components of Bernoulli's equation \eqref{eq:main1} in terms of the free-surface variables \eqref{eq:Tmapvariables}. Starting by differentiating $Y(\xi,t)=\zeta(x,t)$, we find $\zeta_t= Y_t-X_t\zeta_{x}$, $\zeta_x = Y_{\xi}/X_{\xi}$, and $ \zeta_{xx}=(X_{\xi}Y_{\xi \xi}-Y_{\xi}X_{\xi \xi})/X_{\xi}^3$,
of which the last two yield an expression for the curvature, $\kappa = \zeta_{xx}/(1+\zeta_x^2)^{3/2}=(X_{\xi} Y_{\xi \xi} - Y_{\xi} X_{\xi \xi})/J^{3/2}$.
Next, by using the chain rule on $\Phi_{\xi}$ and $\Psi_{\xi}$ and the Cauchy-Riemann equations $\psi_x=-\phi_y$ and $\psi_y=\phi_x$, we find the two equations $\Phi_{\xi}=X_{\xi}\phi_x + Y_{\xi} \phi_y$ and $\Psi_{\xi}=Y_{\xi} \phi_x - X_{\xi} \phi_y$, which may be solved to find
\begin{equation}\label{eq:Tmapphixphiy}
\phi_x=\frac{X_{\xi}\Phi_{\xi}+Y_{\xi}\Psi_{\xi}}{J} \qquad \text{and} \qquad  \phi_y=\frac{Y_{\xi}\Phi_{\xi}-X_{\xi}\Psi_{\xi}}{J}.
\end{equation}
It remains to find expressions for $\phi_t$ and $\phi_{yy}$. We use the chain rule to find $\Phi_{\xi \xi}=X_{\xi \xi}\phi_x+Y_{\xi \xi}\phi_y+(Y_{\xi}^2-X_{\xi}^2)\phi_{yy}+2X_{\xi}Y_{\xi}\phi_{xy}$ and $\Psi_{\xi \xi}=Y_{\xi \xi}\phi_x-X_{\xi \xi}\phi_y+(Y_{\xi}^2-X_{\xi}^2)\phi_{xy}-2X_{\xi}Y_{\xi}\phi_{yy}$, where $\phi_{xx}$ has been eliminated from Laplace's equation \eqref{eq:main3}. We then eliminate $\phi_{xy}$, and substitute for $\phi_x$ and $\phi_y$ from \eqref{eq:Tmapphixphiy}, which yields
\begin{equation}\label{eq:Tmapphiyy}
\begin{aligned}
\phi_{yy}&=\frac{(Y_{\xi}^2-X_{\xi}^2)}{J^2}\Phi_{\xi \xi}+\frac{X_{\xi}X_{\xi \xi}(X_{\xi}^2-3Y_{\xi}^2)+Y_{\xi}Y_{\xi \xi}(3X_{\xi}^2-Y_{\xi}^2)}{J^3}\Phi_{\xi}\\
&\quad -\frac{2X_{\xi}Y_{\xi}}{J^2}\Psi_{\xi \xi}
+\frac{Y_{\xi}X_{\xi \xi}(3X_{\xi}^2-Y_{\xi}^2)+X_{\xi}Y_{\xi\xi}(3Y_{\xi}^2-X_{\xi}^2)}{J^3}\Psi_{\xi}.
\end{aligned}
\end{equation}
The expression \eqref{eq:Tmapphiyy} for $\phi_{yy}$ in conformal variables is required due to the viscous term in Bernoulli's equation \eqref{eq:main1}.
Lastly, we differentiate $\Phi(\xi,t)$ using the chain rule to find
\begin{equation}\label{eq:tmapphit}
\phi_t=\Phi_t-\phi_x X_t -\phi_y Y_t.
\end{equation}
However, substitution of \eqref{eq:tmapphit} into Bernoulli's equation results in components given by $\Phi_t$, $X_t$, and $Y_t$. The last two of these can be specified explicitly through consideration of the kinematic boundary condition.

Substitution of the conformal expressions for $\zeta_t$, $\zeta_{x}$, $\zeta_{xx}$, $\phi_x$, and $\phi_y$ into the kinematic boundary condition yields
\begin{subequations}\label{eq:Tmapkinsubs}
\begin{equation}\label{eq:tmapkinsubsa}
X_{\xi}Y_{t}-Y_{\xi}X_t=Y_{\xi}-\Psi_{\xi} + \frac{2}{Re}\frac{X_{\xi}Y_{\xi \xi}-Y_{\xi}X_{\xi \xi}}{X^2_{\xi}}.
\end{equation}
Furthermore, through the study of the analytic function $z(\xi,\eta,t)=x(\xi,\eta,t)+\i y(\xi,\eta,t)$, we have $\text{Im}[z_t/z_{\xi}]_{\eta=0}=(X_{\xi}Y_t-Y_{\xi}X_t)/J$, for which the numerator is the left-hand side of \eqref{eq:tmapkinsubsa} above. In noting that $\text{Re}[z_t/z_{\xi}]_{\eta=0}=(X_{\xi}X_t+Y_{\xi}Y_t)/J$, we may then use the harmonic relation between the real and imaginary parts of $z_t/z_{\xi}$, $\text{Re}[z_t/z_{\xi}]=-\mathscr{H}[\text{Im}[z_t/z_{\xi}]]$, to find
\begin{equation}\label{eq:tmapkinsubsb}
X_{\xi}X_{t}+Y_{\xi}Y_t=-J \mathscr{H} \bigg[\frac{Y_{\xi}-\Psi_{\xi}}{J} +\frac{2}{Re}\frac{X_{\xi}Y_{\xi \xi}-Y_{\xi}X_{\xi\xi}}{X_{\xi}^2 J}\bigg].
\end{equation}
\end{subequations}

From equations \eqref{eq:tmapkinsubsa} and \eqref{eq:tmapkinsubsb}, we may eliminate $X_t$ to find our time-evolution equation for $Y$, which is given by
\begin{subequations}\label{eq:TmapevolvebothApp}
\begin{equation}\label{eq:TmapyevolveAppA}
Y_t=\frac{X_{\xi}(Y_{\xi}-\Psi_{\xi})}{J}+\frac{2}{Re}\frac{X_{\xi} Y_{\xi \xi} - Y_{\xi} X_{\xi \xi}}{X_{\xi}J}-Y_{\xi} \mathscr{H} \bigg[\frac{Y_{\xi}-\Psi_{\xi}}{J} +\frac{2}{Re}\frac{X_{\xi}Y_{\xi \xi}-Y_{\xi}X_{\xi\xi}}{X_{\xi}^2 J}\bigg],
\end{equation}
Next, we substitute the above expressions into the dynamic boundary condition to find the following time evolution equation for $\Phi$,
\begin{equation}\label{eq:TmapphievolveAppB}
\begin{aligned}
\Phi_t=&\frac{\Psi_{\xi}^2-\Phi_{\xi}^2}{2J} - \frac{Y}{F^2}+\frac{P}{F^2}\frac{Y_{\xi}}{X_{\xi}} + \frac{B \kappa}{F^2}+\frac{X_{\xi}\Phi_{\xi}}{J}+\Phi_{\xi}\mathscr{H} \bigg[\frac{\Psi_{\xi}-Y_{\xi}}{J} -\frac{2}{Re}\frac{X_{\xi}Y_{\xi \xi}-Y_{\xi}X_{\xi\xi}}{X_{\xi}^2 J}\bigg] \\
&+\frac{2}{Re}\bigg[\frac{(Y_{\xi} X_{\xi \xi} - X_{\xi} Y_{\xi \xi})}{X^2_{\xi}J}\Psi_{\xi}
+\frac{Y_{\xi}X_{\xi \xi}(Y_{\xi}^2-3X_{\xi}^2)+X_{\xi}Y_{\xi\xi}(X_{\xi}^2-3Y_{\xi}^2)}{J^3}\Psi_{\xi}\\
&+\frac{X_{\xi}X_{\xi \xi}(3Y_{\xi}^2-X_{\xi}^2)+Y_{\xi}Y_{\xi \xi}(Y_{\xi}^2-3X_{\xi}^2)}{J^3}\Phi_{\xi} +\frac{X_{\xi}^2-Y_{\xi}^2}{J^2}\Phi_{\xi \xi}+\frac{2X_{\xi}Y_{\xi}}{J^2}\Psi_{\xi \xi}\bigg],
\end{aligned}
\end{equation}
\end{subequations}
Combined with the harmonic relations $X_{\xi}=1-\mathscr{H}[Y_{\xi}]$ and $\Psi_{\xi}=\mathscr{H}[\Phi_{\xi}]$, equations \eqref{eq:TmapevolvebothApp} form our evolution equations for the free-surface variables in the moving frame. 

\subsection{The numerical method for time evolution}\label{sec:unsteadymethod}
Our numerical implementation of the time-evolution equations \eqref{eq:TmapyevolveAppA} and \eqref{eq:TmapphievolveAppB} is similar to that used by \cite{milewski2010dynamics} for solitary waves, and \cite{shelton2023structure} for standing waves. We now provide details on the implementation of this method. We note that other methods exist to evolve surface water waves in time such as the graph-based formulation from \cite{wilkening2012overdetermined}, which used Dirichlet-to-Neumann operators to evaluate the kinematic and dynamic boundary conditions in order to study steep standing waves both with and without capillarity.

We begin at $t=0$ with a specified initial condition for $Y(\xi,0)$, $\Phi(\xi,0)$, $B$, $F$, $Re$, and $P$, for which $X(\xi,0)$ and $\Psi(\xi,0)$ are computed spectrally from harmonic relations \eqref{eq:harmonicrelationsC} and \eqref{eq:harmonicrelationsD}. The number of grid points in the initial condition is typically $N=1024$. Note that the steady solutions from \S\,{\ref{sec:steadyresults}} only yield $\Phi_{\xi}(\xi,0)$. We determine $\Phi(\xi,0)$ from this by integrating spectrally with the Fourier multiplier $1/(2 \pi \mathrm{i} k)$ if $k\neq0$ and $0$ if $k=0$. The constant of integration is subsequently determined from condition \eqref{eq:phicond}, which we enforce in Fourier space by setting the constant level of $\Phi X_{\xi}$ to be zero.

The temporal step size, $\Delta t = t_{j+1}-t_{j}$, is typically specified throughout the evolution process as $\Delta t =0.00015$, but this may need to be smaller for larger values of $N$. We then use the fourth-order Runge-Kutta method to obtain solution values at the next time step, $Y(\xi,t_{j+1})$ and $\Phi(\xi,t_{j+1})$. This requires the evaluation of evolution equations \eqref{eq:TmapyevolveAppA} and \eqref{eq:TmapphievolveAppB}.
Derivatives and Hilbert transforms are computed spectrally in Fourier space (see \S\,{\ref{sec:numericalsteady}} for more details of this). Nonlinear terms are computed by multiplication in real space, during which aliasing errors are reduced by padding each solution with an additional $N/2$ Fourier modes, which are subsequently set to zero.

Note that energy and momentum are not conserved quantities of this viscous formulation. Across our simulations in \S\,{\ref{sec:nonlinearstability}}, which were each ran to $t=1000$, the mass $M=\int_{-1/2}^{1/2}YX_{\xi}\mathrm{d}\xi$ is conserved to within $10^{-13}$.

\section{Energy evolution due to viscosity}\label{App:energydecay}
In this section, we directly calculate the energy decay induced by the damping in the kinematic and dynamic boundary conditions \eqref{eq:intro}. In the absence of viscous dissipation and wind forcing, the nondimensional bulk energy,
\begin{equation}\label{eq:Appenergy}
E=\int_{-1/2}^{1/2}\int_{-\infty}^{\zeta} \frac{F^2}{2} \lvert \nabla \phi \rvert^2 \mathrm{d}y \mathrm{d}x +\int_{-1/2}^{1/2} \bigg[B\big((1+\zeta_x^2)^{1/2}-1\big)+\frac{\zeta^2}{2}\bigg]\mathrm{d}x,
\end{equation}
is a Hamiltonian of the system \citep{zakharov1968stability}. To derive \eqref{eq:Appenergy}, we nondimensionalised each component of the dimensional energy $\hat{E}=(\rho/2) \int_{-\lambda/2}^{\lambda/2} \int_{-\infty}^{\hat{\zeta}} \lvert \nabla \hat{\phi}\rvert^2 \mathrm{d}\hat{y}\mathrm{d}\hat{x}+ \int_{-\lambda/2}^{\lambda/2}[\sigma ({[1+\hat{\zeta}_{\hat{x}}^2]^{1/2}}-1) + g \rho \hat{\zeta}^2/2]\mathrm{d}\hat{x}$, and defined $\hat{E}=\rho g \lambda^3 E$. Note that our previous energy expression \eqref{eq:energy} may be derived by applying the divergence theorem to the double integral in \eqref{eq:Appenergy}, and writing in terms of conformal variables. Differentiating \eqref{eq:Appenergy} with respect to time yields
\begin{equation}\label{eq:Appenergy2}
\frac{\mathrm{d}E}{\mathrm{d}t}=\int_{-1/2}^{1/2}\int_{-\infty}^{\zeta}F^2 \nabla \phi \cdot \nabla \phi_{t} \mathrm{d}y \mathrm{d}x +\int_{-1/2}^{1/2} \bigg[\frac{F^2\lvert \nabla \phi \rvert^2 \zeta_{t}}{2}+\frac{B \zeta_{x}\zeta_{xt}}{(1+\zeta_x^2)^{1/2}}+\zeta \zeta_{t}\bigg]\mathrm{d}x.
\end{equation}
We may now apply the divergence theorem to the first integral above by writing $\nabla \phi \cdot \nabla \phi_{t}=\nabla \cdot (\phi_{t}\nabla \phi)-\phi_{t}\nabla^2 \phi$, which, together with integration by parts on the surface tension term yields
\begin{equation}
\label{eq:Appenergy3}
\begin{aligned}
\frac{\mathrm{d}E}{\mathrm{d}t} &= \int_{-1/2}^{1/2} F^2 \phi_{t} \phi_{n} \frac{\mathrm{d}s}{\mathrm{d}x} \mathrm{d}x +\int_{-1/2}^{1/2} \bigg[\frac{F^2\lvert \nabla \phi \rvert^2}{2}-\frac{B \zeta_{xx}}{(1+\zeta_x^2)^{3/2}} + \zeta \bigg]\zeta_{t}\mathrm{d}x,\\
&=-\frac{F^2}{Re} \int_{-1/2}^{1/2} \left(\bigg[-\frac{1}{2}\lvert \nabla \phi \rvert^2+\frac{B \zeta_{xx}}{F^2(1+\zeta_x^2)^{3/2}} - \frac{\zeta}{F^2} \bigg] \zeta_{xx} - \phi_{xx}\phi_n \frac{\mathrm{d}s}{\mathrm{d}x} \right) \mathrm{d}x,\\
&=-\frac{F^2}{Re} \bigg( \int_{-1/2}^{1/2} \left[-\frac{1}{2}\lvert \nabla \phi \rvert^2 \zeta_{xx} +\frac{B \zeta_{xx}^2}{F^2(1+\zeta_x^2)^{3/2}} + \frac{\zeta_x^2}{F^2} \right] \mathrm{d}x + \int_{-1/2}^{1/2}\int_{-\infty}^{\zeta} \lvert \nabla \phi_x \rvert^2 \mathrm{d}y \mathrm{d}x \bigg),\\
\end{aligned}
\end{equation}
where $\phi_{n}$ is the normal derivative of $\phi$ and $s$ is arclength. In the second line we have used the boundary conditions (\ref{eq:intro}), noting that the kinematic boundary condition can be written $\zeta_t  = \phi_n \frac{\mathrm{d}s}{\mathrm{d}x} + \frac{2}{Re} \zeta_{xx}$. In the third line we have applied the divergence theorem and integrated two terms by parts. We see from this that the energy evolution is inversely proportional to $Re$ and that the linear dissipation model in the nonlinear system leads to one higher-order sign indefinite term $\frac{1}{2}\lvert \nabla \phi \rvert^2 \zeta_{xx}$. We have been unable to identify any situation in which the integral of this sign indefinite term dominates the other components of \eqref{eq:Appenergy3} and leads to $\mathrm{d}E/ \mathrm{d}t \geq 0$. In all of our numerical simulations without wind forcing, the energy has been a decreasing function of time.

\bibliographystyle{jfm}
%\bibliography{short}

\providecommand{\noopsort}[1]{}

\end{document}